\documentclass[english]{nature}
\usepackage[colorlinks=true,citecolor=blue,linkcolor=magenta]{hyperref}
\usepackage{lineno}
\usepackage{amsmath}
\usepackage{amsfonts}
\usepackage{stmaryrd}
\usepackage{amssymb}
\usepackage{graphicx}
\usepackage{dcolumn}
\usepackage{bm}
\usepackage[font={sf, small},labelfont=bf]{caption}
\usepackage{upgreek}

\usepackage{soul}
\usepackage{changes}

\begin{document}

\title{Hybrid Integrated Photonics Using Bulk Acoustic Resonators}

\author{\noindent Hao Tian$^{1}$, Junqiu Liu$^{2}$, Bin Dong$^{1}$, J Connor Skehan$^{2}$, Michael Zervas$^{2}$, Tobias J. Kippenberg$^{2,\dagger}$, Sunil A. Bhave$^{1,\dagger}$}
\maketitle

\begin{affiliations}
\item OxideMEMS Lab, Purdue University, 47907 West Lafayette, IN, USA
\item Institute of Physics, Swiss Federal Institute of Technology Lausanne (EPFL), 1015 Lausanne, Switzerland

\normalsize{$^\dagger$ E-mail: tobias.kippenberg@epfl.ch, bhave@purdue.edu}
\end{affiliations}

\begin{abstract}
Integrated photonic devices based on Silicon Nitride (Si$_3$N$_4$) waveguides allow for the exploitation of nonlinear frequency conversion, exhibit low propagation loss, and have led to advances in compact atomic clocks, supercontinuum generation, optical sensing, ultrafast ranging, coherent communication, and spectroscopy. Yet, in contrast to silicon, the amorphous nature of (Si$_3$N$_4$) requires combination with other materials to achieve active tuning or modulation. Approaches so far range from thermal tuning, integration with Graphene and atomically thin semiconductors to solution based processing of piezoelectric materials. 
Here, microwave frequency acousto-optic modulation is realized by exciting high overtone bulk acoustic wave resonances (HBAR resonances) in the photonic stack. These confined mechanical stress waves exhibit vertically transmitting, high quality factor (Q) acoustic Fabry-Perot resonances that extend into the Gigahertz domain, and offer stress-optical interaction with the optical modes of the microresonator. Although HBAR are ubiquitously used in modern communication, and often exploited in superconducting circuits, this is the first time they have been incorporated on a photonic circuit based chip.
The electro-acousto-optical interaction observed within the optical modes
exhibits high actuation linearity, low actuation power and negligible crosstalk. Using the electro-acousto-optic interaction, fast optical resonance tuning is achieved with sub-nanosecond transduction time. By removing the silicon backreflection, 
broadband acoustic modulation at 4.1 and 8.7 GHz is realized with a 3 dB bandwidth of 250 MHz each. The novel hybrid HBAR nanophotonic platform demonstrated here, allowing on chip integration of micron-scale acoustic and photonic resonators, can find immediate applications in tunable microwave photonics, high bandwidth soliton microcomb stabilization, compact opto-electronic oscillators, and in microwave to optical conversion schemes. Moreover the hybrid platform allows implementation of momentum biasing, which allows realization of on chip non-reciprocal devices such as isolators or circulators and topological photonic bandstructures.
\end{abstract}

 
\setlength{\parskip}{12pt}%

Integrated photonics has drawn increasing attention in recent years. While within the last decade, silicon photonics has transitioned from laboratory based research into commercial products used in data-centers, major efforts are still underway with regards to silicon nitride (Si$_3$N$_4$) photonic devices \cite{Moss:13}. 
The platform has attracted intense efforts due to its wide transparency window from the visible to mid-infrared, ultralow linear losses, absence of two photon absortion in the telecommunication band, space-compatibility \cite{brasch2014radiation}, large Kerr nonlinearity ($\chi ^{(3)}$), and wide geometric flexibility for waveguide dispersion engineering \cite{Gaeta:19}. The material has been the material of choice for soliton microcomb generation \cite{Kippenberg:11}, and supercontinuum generation \cite{Gaeta:19}, optical filters \cite{huffman2017}, gyroscopes \cite{gundavarapu2018}, and optical interconnects \cite{yao2018} have been demonstrated. Recent advances of Si$_3$N$_4$-based dissipative Kerr soliton (DKS) microcombs, have included octave-spanning comb spectra \cite{Li:17, Pfeiffer:17}, ultralow initiation power \cite{Liu:18a, Stern:18, Raja:19}, and microcomb repetition rates in the microwave domain \cite{Liu:19}. 

Reliable and fast resonance tuning of Si$_3$N$_4$ microring resonators is becoming an important asset and requirement for a number of applications in integrated nonlinear photonics. For example, high bandwith tuning allows microcomb repitition rate stabilization \cite{spencer2018}, or resonance tuning for filters. Likewise, recently emerged platforms such as spatio-temporal modulation based optical non-reciprocity \cite{shi2018} and topological optical bandstructures both require GHz speed modulation within optical microresonators, which poses stringent requirements on the cross-talk and size.

However, due to the inversion symmetry, and thus the lack of $\chi ^{(2)}$ nonlinearity, it is difficult to electrically modulate the refractive index of Si$_3$N$_4$. Traditionally, the thermo-optical effect is employed to fulfill the tuning requirement \cite{Xue:16, Joshi:16} which, however, presents low tuning speed ($\sim$1 ms), high power consumption ($\sim$1 mW), and thermal cross-talk. These drawbacks make it incompatible with large-scale integration and cryogenic applications \cite{Moille:19}. Although hybrid integration with various electro-optical materials, e.g., graphene \cite{yao2018gate}, lead zirconate titanate (PZT)  \cite{alexander2018}, lithium niobate (LiNbO$_3$) \cite{ahmed2019}, and monolayer WS$_2$ \cite{datta2019low} has made significant progresses, there are still remaining challenges related to CMOS-compatibility, fabrication complexity, optical losses, and dispersion engineering. To fully utilize the maturity and advantages of Si$_3$N$_4$ photonics, new tuning mechanisms which retain the original optical properties are needed. 

\begin{figure}[t]
\centering\includegraphics[width=\textwidth]{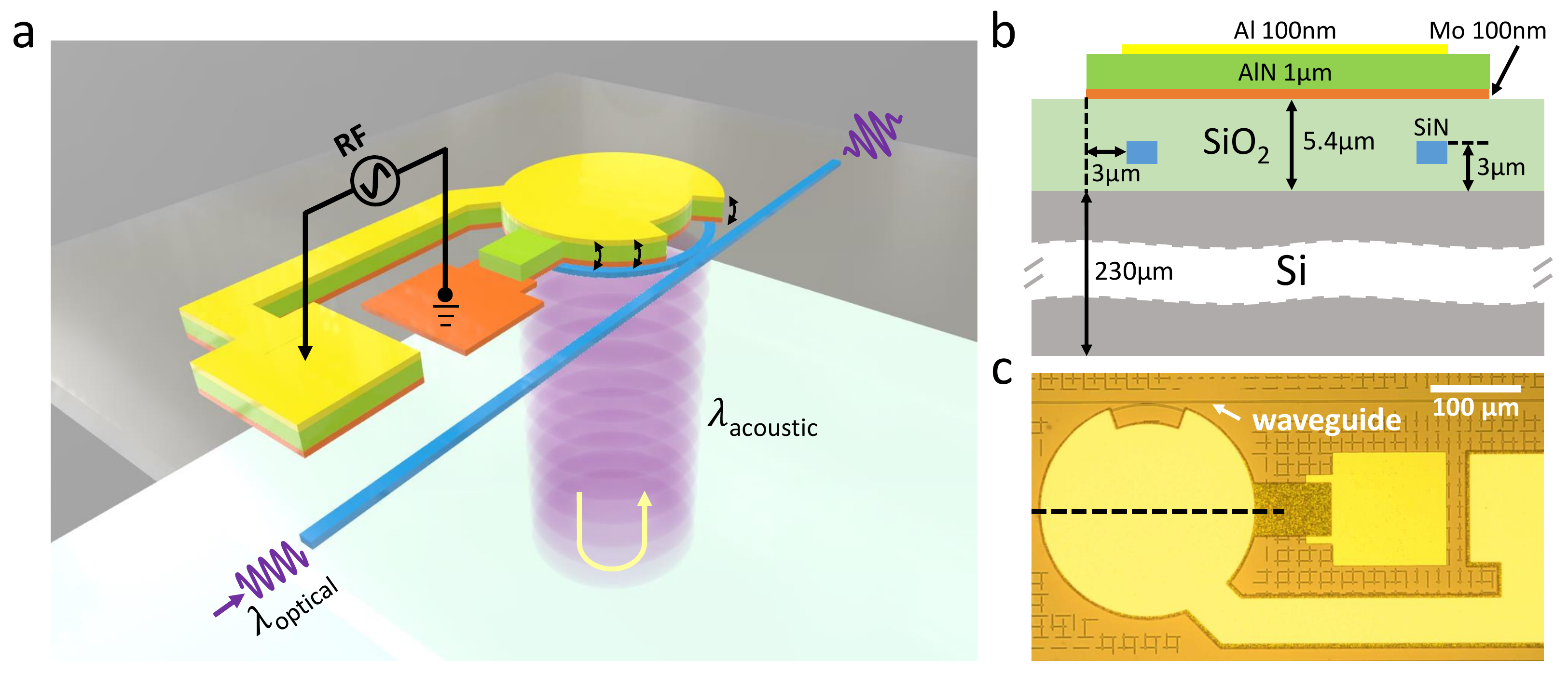}
\caption{\textbf{Hybrid nanophotonic high-overtone bulk acoustic resonator (HBAR) platform}. \textbf{a} 3D schematic illustrating excitation of bulk acoustic wave resonances via a macroscopic piezoelectric actuator, which transmit vertically into the stack and form acoustic standing waves inside the various acoustic Fabry-Perot cavities. The resonance enhanced mechanical stress changes the waveguide's effective index via the stress-optical effect, and thereby modulate the output optical intensity. \textbf{b} Cross-section of the device along the black dashed line in \textbf{c}, with critical dimensions labeled. The AlN piezoelectric actuator is placed directly on top of the Si$_3$N$_4$ microring resonator which is embedded in 5.4-$\upmu$m of SiO$_2$ cladding. \textbf{c} Optical microscope image of the fabricated device. }
\label{Fig:figure1}
\end{figure}

The stress-optical effect, discovered over a hundred years ago, has recently gained attention for its role in the modulation of Si$_3$N$_4$ waveguides and microring resonators both theoretically\cite{huang2003, van2019} and experimentally \cite{hosseini2015, jin2018piezoelectrically}, thanks to the advances in Micro-Electro-Mechanical Systems (MEMS) \cite{midolo2018nano}. In this work, by integrating aluminium nitride (AlN) piezoelectric actuators onto Si$_3$N$_4$ photonic devices, we demonstrate, to the best of our knowledge, the first acousto-optic modulation of Si$_3$N$_4$ microring resonators using High-overtone Bulk Acoustic wave Resonances (HBAR) \cite{macquarrie2013}. Sub-micron wavelength acoustic waves are excited by macroscopic actuators, which transmit vertically into the substrate and perpendicular to optical paths (see Fig. \ref{Fig:figure1}a). Trapped inside a Fabry-Perot-like acoustic cavity that is formed by the top and bottom surfaces of the entire substrate, a rich family of acoustic resonant modes is efficiently excited at the microwave frequencies up to 6 GHz. The coupling between vertical acoustic waves and in-plane optical circuits makes it possible for independent optimization of the actuator and optical components. The high lateral acoustic mode confinement further enables low cross-talk and compact integration. These features are in stark contrast with conventional surface acoustic wave (SAW) acousto-optic modulation (AOM) \cite{tadesse2014,van2019} which requires an inter-digital transducer (IDT) with sub-micron electrode fingers, coplanar integration of IDT and optical waveguides, and perfect termination of acoustic waves to reduce cross-talk. 

The same structure platform is capable of linear bi-directional tuning of Si$_3$N$_4$ microring resonators, which shows a high power efficiency of 5 nW per picometer resonant wavelength tuning (compared to thermal tuning on the order of 1 mW/pm \cite{Xue:16}). Compared with PZT \cite{hosseini2015, jin2018piezoelectrically}, AlN has the advantage of of high actuation linearity, no hysteresis, high power handling (breakdown field $>$100 V/$\upmu$m), and low current leakage. Here, the tuning speed is fully explored, demonstrating sub-nanosecond switching capability, which is mainly limited by intrinsic acoustic resonances. Broadband (250 MHz) AOMs around 4.1 and 8.7 GHz are realized by damping acoustic resonances inside the Si substrate. The simple but efficient stress-optical platform demonstrated in this work may find widespread applications in Si$_3$N$_4$ microwave photonics \cite{marpaung2019}, such as the repetition rate stabilization and tuning of soliton microcombs via injection-locking \cite{obrzud2017}, on-chip optomechanical frequency comb generation \cite{fan2019}, and comb-assisted microwave photonic filters \cite{supradeepa2012}.

As a well-known piezoelectric thin film, c-axis oriented polycrystalline AlN (with a piezoelectric coefficient $e_{33}\sim$1.55 C/m$^2$ \cite{tsubouchi1985}) is utilized to form the piezoelectric actuator which will be deformed and generate stress around the Si$_3$N$_4$ waveguide when a vertical electric field is applied. As shown in Fig. \ref{Fig:figure1}a, a disk-shaped actuator is placed directly on top of the Si$_3$N$_4$ ring resonator, with the microring resonator positioned at the outer edge of the disk actuator. As we drive the actuator harmonically, vibration of the AlN disk will launch an acoustic wave vertically into the substrate (Fig. \ref{Fig:figure1}a). Since the bottom surface of the substrate is smooth and flat, the acoustic wave will be reflected and subsequently bounce back and forth between the top and bottom surfaces. Working as an acoustic Fabry-Perot cavity, the counter-propagating acoustic waves will constructively interfere when the cavity length is an integer number of the acoustic wavelength, with acoustic energy trapped inside the cavity. These bulk acoustic standing waves will enhance the stress field around optical waveguides and modulate the effective index through stress-optical effect \cite{huang2003}. 

The stack's cross-section is illustrated in higher detail in Fig. \ref{Fig:figure1}b. To apply the vertical electric field, an AlN film of 1 $\upmu$m thickness is sandwiched between top Al (100 nm thickness) and bottom Mo (100 nm thickness) metal layers. The Si$_3$N$_4$ waveguides are fabricated using a subtractive process \cite{Luke:13}, having a height of 800 nm and width of 1.8 $\upmu$m, and which are fully buried in SiO$_2$ cladding 5.4 $\upmu$m thick (2.4 $\upmu$m SiO$_2$ from the Mo layer to prevent metal absorption). The entire device sits on a 230-$\upmu$m-thick Si substrate. The radius of the microring resonator is 118 $\upmu$m, and is placed 3 $\upmu$m within the edge of disk actuator. The final fabricated device is shown in Fig. \ref{Fig:figure1}c (see Methods for more fabrication details). The area around waveguide-to-microring coupling region is opened to prevent any actuation-induced perturbation of light coupling between bus waveguide and microring resonator.   

\begin{figure}[htbp]
\centering\includegraphics[width=\textwidth]{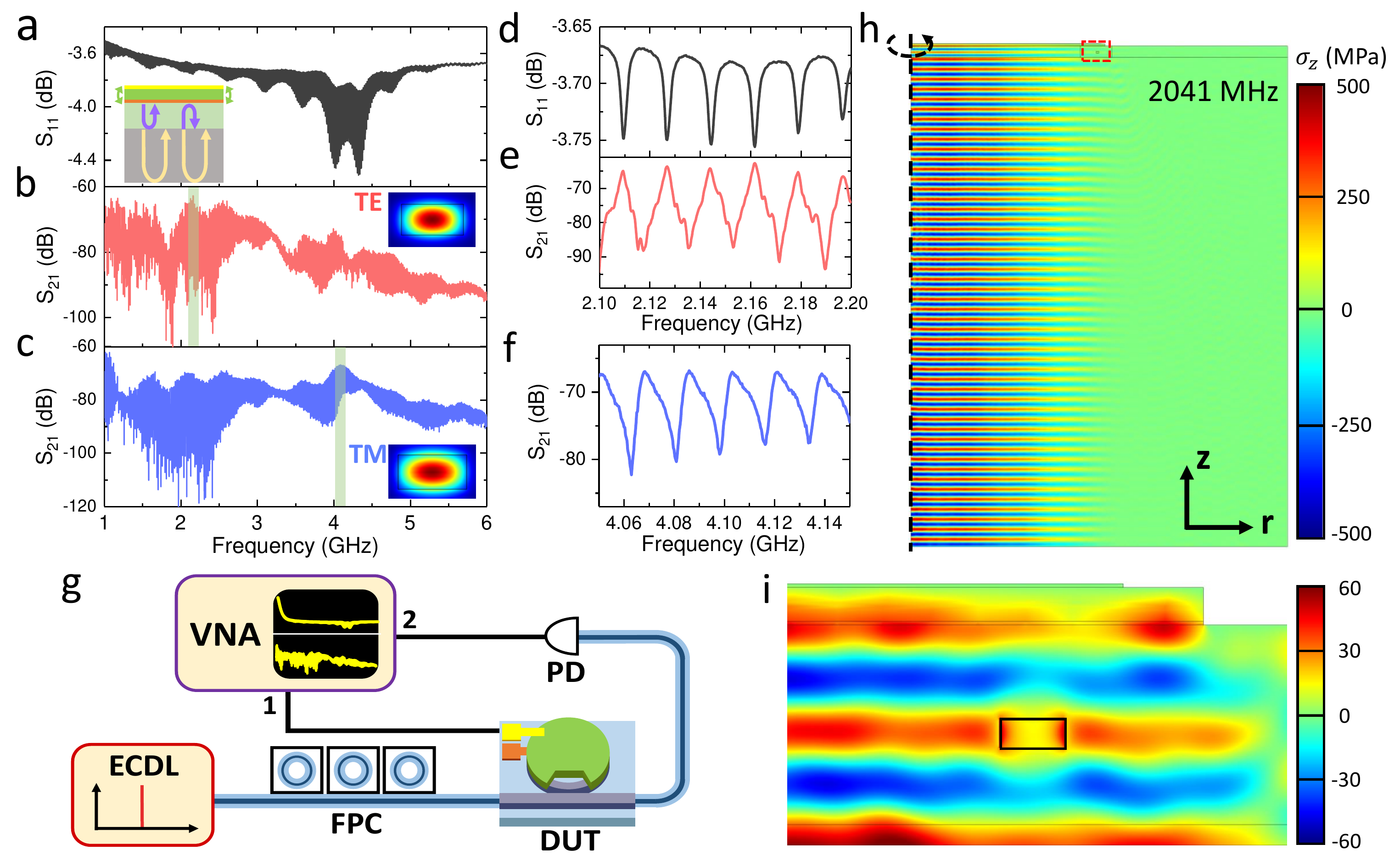}
\caption{\textbf{Microwave frequency electro-acousto-optic modulation.} \textbf{a} Electromechanical S$_{11}$ spectrum from 1 to 6 GHz. A range of equidistant bulk acoustic resonances is found to exist over a broad frequency range. The inset schematic illustrates the acoustic wave reflection at interfaces. \textbf{b}, \textbf{c} Optomechanical S$_{21}$ responses of TE and TM modes demonstrate acousto-optic modulation covering multiple octave-spanning microwave frequencies. Due to different optical mode shapes (insets in \textbf{b} and \textbf{c}) and thus acousto-optic mode overlap, TE and TM modes show dissimilar S$_{21}$ spectra. \textbf{d}, \textbf{e} The zoom-in of S$_{11}$, and TE mode's S$_{21}$ within the window (green shaded area in \textbf{b}) around 2 GHz. \textbf{f} The zoom-in of TM mode's S$_{21}$ around 4 GHz in \textbf{c}. The resonances distribute evenly with an FSR of 17.5 MHz. \textbf{g} Schematic of the setup for measuring electromechanical and optomechanical response. ECDL: external cavity diode laser, PC: polarization controller, DUT: device under test, PD: photo-diode, VNA: vector network analyzer. \textbf{h} Numerical simulation of vertical stress $\sigma _z$ distribution for one typical acoustic resonant mode at 2.041 GHz under 1 V driving field, with a zoom-in around the optical waveguide (red box in \textbf{h}) shown in \textbf{i}. At several GHz, the acoustic wavelength is similar in scale to optical wavelength and waveguide structure.}
\label{Fig:figure2}
\end{figure}

Experimentally, the electromechanical reflection parameter S$_{11}$ is first measured using port 1 of a network analyzer (Fig. \ref{Fig:figure2}g), which characterizes the energy conversion from electrical to mechanical vibration as a dip. As shown in Fig. \ref{Fig:figure2}a, a series of resonance dips is found to evenly distribute over multiple octaves in the microwave regime. A zoom-in of the spectrum is illustrated in Fig. \ref{Fig:figure2}d which highlights the resonance shape, linewidth ($\sim$3 MHz), and acoustic free spectral range (FSR) of around 17.5 MHz. The narrow linewidth demonstrates high mechanical Q ($\sim$1000) which is mainly limited by the intrinsic acoustic loss in the substrate and scattering at interfaces. It is interesting to note that the envelope of these sharp resonances varies slowly and smoothly with a period of $\sim$ 490 MHz. This is caused by the resonance inside the 5.4 $\upmu$m SiO$_2$ cladding layer, which stems from acoustic wave reflections at the Si-SiO$_2$ interface due to an acoustic impedance mismatch. The position of SiO$_2$ resonance is located at the node of the envelope, whereas anti-resonance is located at the anti-node. Intuitively speaking, at the anti-resonance of the SiO$_2$ layer, it works as an acoustic anti-reflection coating such that more acoustic energy will transmit into the Si substrate, which has larger acoustic impedance, and thus better electromechanical conversion. When the acoustic half wavelength matches the 1 $\upmu$m AlN thickness, the AlN layer reaches its fundamental resonance mode and becomes more efficient in excitation of acoustic waves around 4 -- 4.3 GHz. It can be seen that, by optimizing AlN and SiO$_2$ thicknesses, the resonances of these two cavities can be misaligned to further improve acoustic wave excitation. Additionally, the coupling between the Si substrate cavity and the SiO$_2$ and AlN cavities causes periodic fluctuations of the FSR and higher order dispersion, as will be discussed later. 

The acousto-optic modulation of the microring resonator can be characterized by an optomechanical S$_{21}$ measurement as shown in Fig. \ref{Fig:figure2}g, where by biasing the input laser ($\sim$1550 nm) at the slope of the optical resonance, its output intensity is modulated as we launch a -5 dBm RF signal from port 1 of the VNA. The intensity modulation of the optical signal is measured using a photodiode and sent back to port 2 for the S$_{21}$ measurement. Note that no optical and electrical amplifiers are employed in an effort to preserve direct electro-opto-mechanical transduction. The optomechanical S$_{21}$ measurements are performed for both the transverse electric (TE) and transverse magnetic (TM) optical modes, as shown in Fig. \ref{Fig:figure2}b and c respectively. As expected, a broad range of periodic peaks is observed which correspond to each HBAR mode in S$_{11}$. Due to the different optical mode profiles of the TE and TM modes (and thus acousto-optic mode overlaps) and optical Q factors, they respond differently, with the TE mode response strongest around 2 GHz and the TM mode responding most strongly around 4 GHz. Since the TE mode shows higher optical Q, it will enter the resolved sideband regime at frequencies far beyond its linewidth (1 -- 2 GHz), where the modulation sidebands are suppressed when biasing at the resonance slope. The zoom-in of highlighted regions (green shaded areas) are as shown in Fig. \ref{Fig:figure2}e and f, illustrating clear peaks with high contrast ($>$20 dB) between resonance and anti-resonance. Interestingly, the same measurements are performed for devices with a different actuator lateral shape, which show similar results for S$_{11}$ and S$_{21}$ of the TE and TM modes (see Supplementary Material). This indicates that the HBAR mode distribution and optomechanical spectra are less related to the shape of the actuator, but more to the vertical stack. This relaxes the requirements on actuator shape and size, which lends itself with high design freedom and small footprint.

A numerical (COMSOL) simulation of one typical acoustic mode at 2.041 GHz is shown in Fig. \ref{Fig:figure2}h, and a zoom-in around the optical waveguide is in Fig. \ref{Fig:figure2}i. The acoustic standing wave distributes uniformly over the entire substrate, which indicates that the optical circuits can be buried deeply inside the SiO$_2$ cladding, free from the trade-off between actuation efficiency and absorption losses due to metal as found in traditional optical modulators. Also, an acoustic wavelength that is comparable to optical wavelength and waveguide structure at microwave frequencies is achieved with macroscopic actuators. As is evident, the simplicity of our structure presents advantages such as simple fabrication, high fabrication tolerance, high rigidity, and high power handling. Additionally, it can be seen from Fig. \ref{Fig:figure2}h that the acoustic mode is largely confined beneath the actuator which guarantees low electromechanical cross-talk (-60 dB) between adjacent devices (see Supplementary Material). These features may supplement conventional SAW based AOM for future microwave photonics applications with ultra-high optical Q, low cost, and dense integration.

\begin{figure}[htbp]
\centering\includegraphics[width=13.5 cm]{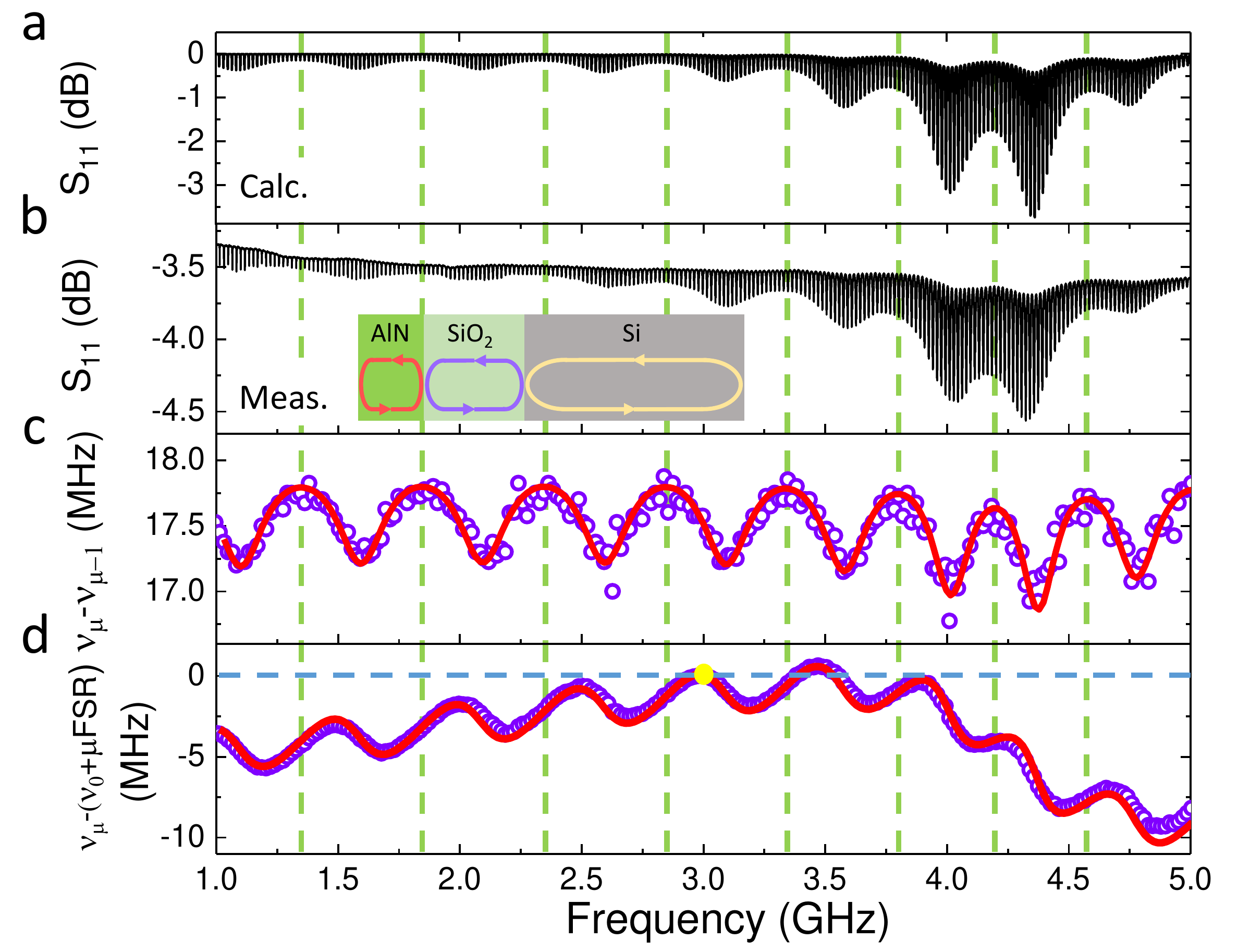}
\caption{\textbf{Mechanical dispersion analysis of HBAR modes.} \textbf{a} Calculated and \textbf{b} measured S$_{11}$ spectrum showing good agreement between the electromechanical model and experiment. Each green dashed line denotes the location of a resonance from the SiO$_2$ cavity. The inset in \textbf{b} illustrates the coupling between Si, SiO$_2$, and AlN cavities. \textbf{c} The measured (circle) and calculated (solid line) frequency difference between each pair of adjacent Si resonances, showing a periodic variation of local FSR around an average value of 17.5 MHz. Note that the maxima of FSR align with green dashed lines where the SiO$_2$ resonances are located. \textbf{d} Measured (circle) and calculated (solid line) higher order dispersion represented by the frequency deviation from an equidistant frequency grid (with average FSR = 17.515 MHz), referencing to mode $\nu _0$ (= 3.0145 GHz, labeled as yellow dot). $\mu$ is the mode number difference relative to the mode at 3.0145 GHz.}
\label{Fig:figure3}
\end{figure}

An analytical electromechanical model is established to help us get deep understanding of the mechanical performance of the device (see Supplementary Material for details). The S$_{11}$ response is first calculated as shown in Fig. \ref{Fig:figure3}a, illustrating high accuracy of the model as compared to the experiment (Fig. \ref{Fig:figure3}b). As mentioned above, the coupling from SiO$_2$ and AlN cavities not only modulates the resonances' magnitude envelope, but also the dispersion of mechanical modes (deviation from equidistant spectrum). Fig. \ref{Fig:figure3}c clearly shows the variation of frequency difference (local FSR) between each pair of adjacent resonances. The local FSR varies nearly periodically around an average value of 17.5 MHz with the same period as the envelope in Fig. \ref{Fig:figure3}b. By comparing Fig. \ref{Fig:figure3}b and c, it can be found that at each node of the envelope, the FSR reaches maximum value, which means the spacing of Si resonances increases near the SiO$_2$ resonance. From the standpoint of the Si cavity, the variation of FSR can be understood intuitively that, the wave reflected back into the Si substrate from the Si-SiO$_2$ interface experiences varied phases relative to the SiO$_2$ resonance. For example, around SiO$_2$ resonance, the Si-SiO$_2$ interface locates at the node of an acoustic stress wave with near-zero stress (maximum displacement), which presents a free boundary condition. Far beyond SiO$_2$ resonances, the interface is at an anti-node corresponding to fixed boundary condition (zero displacement). These various boundary conditions each impose a particular phase for a given reflected wave, and thus change the effective cavity length of Si. 

However, for the AlN and SiO$_2$ cavities, the green dashed lines that denote the location of each SiO$_2$ resonance bunch together around 4 GHz where the AlN cavity resonance is located. Also, it can be seen that the average FSR of acoustic modes decreases around the AlN resonance. Based on these observations and the fact that the acoustic impedance of SiO$_2$ is smaller than Si and then AlN, we can conclude that, for two coupled acoustic cavities, the small cavity (e.g., SiO$_2$) with smaller acoustic impedance tends to decrease (increase) the effective cavity length (FSR) of the big cavity (e.g., Si) when it's on resonance compared to off resonance, and vice versa \cite{zhang2003resonant}. This is, to some extent, similar to coupled optical cavities by treating acoustic impedance as the effective refractive index. 

The higher order dispersion is presented in Fig. \ref{Fig:figure3}d, which shows the frequency deviation of each resonance from the ideally even distribution with reference to mode at 3.0145 GHz and period of 17.515 MHz. Mathematically, it can be interpreted as the integral of Fig. \ref{Fig:figure3}c (that is, the accumulation of FSR deviation relative to 17.515 MHz) with respect to the origin at 3.0145 GHz. The higher order dispersion also shows periodic variation caused by the coupling between the Si and SiO$_2$ cavities, which varies between normal and anomalous group velocity dispersion. The roll-off starting around 3.5 GHz is caused by reduced FSR due to coupling of the AlN cavity. This study of mechanical dispersion could benefit the growing field of mechanical dispersion engineering for future applications (e.g., mechanical solitons) and future devices by optimizing Si, SiO$_2$, and AlN thicknesses or by choosing materials with different acoustic impedance. In this sense, further theoretical and numerical studies are necessary for a more complete understanding of the acoustic wave propagation and mechanical cavity coupling in such a platform. 

\begin{figure}[htbp]
\centering\includegraphics[width=\textwidth]{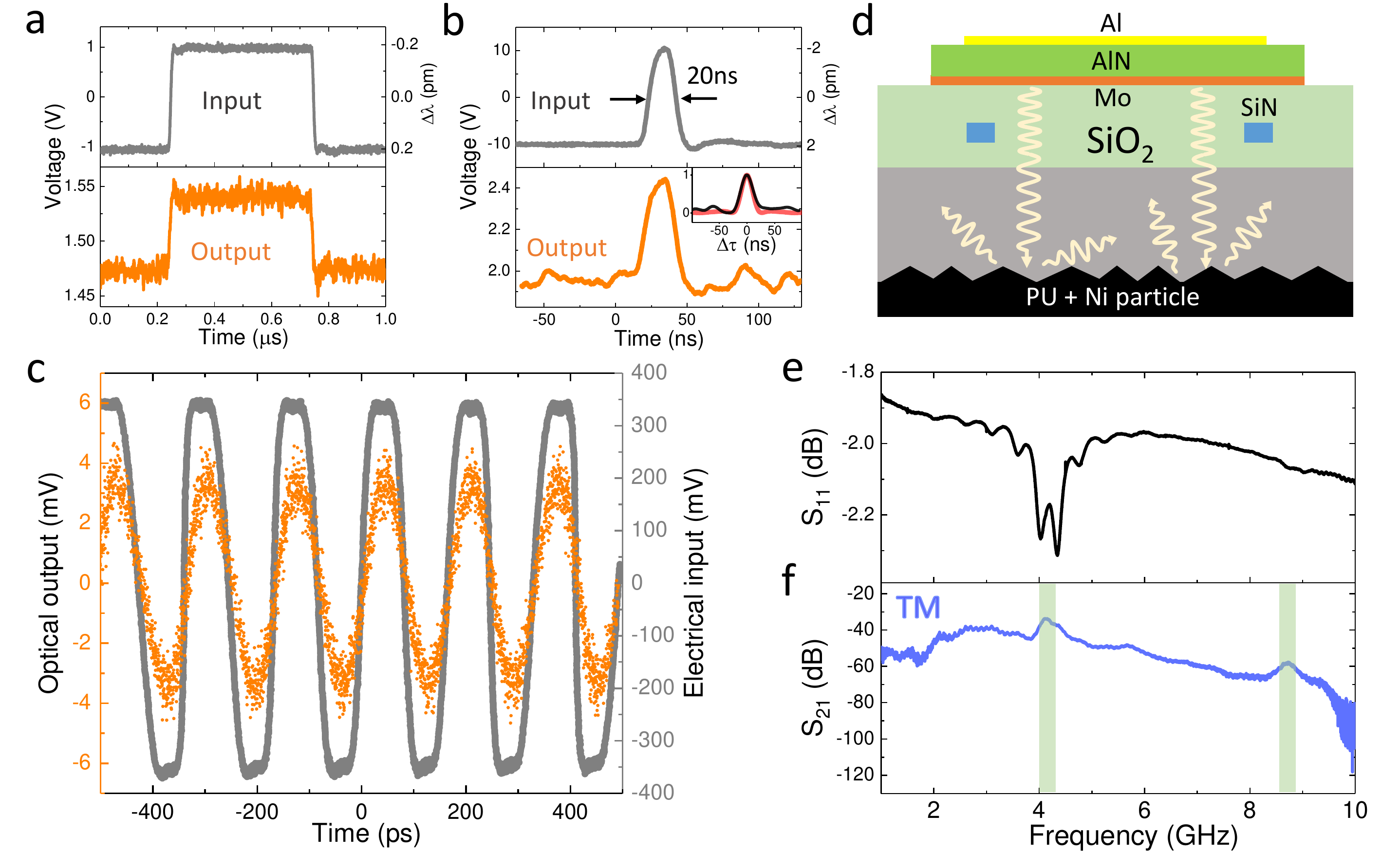}
\caption{\textbf{High speed actuation and broadband electro-acousto-optic modulation.} \textbf{a} Time domain response (orange) to a small-signal square wave (gray) with 2 V V${_\text{pp}}$, 1 MHz repetition rate, and a 50\% duty cycle. The sharp peaks at tuning edges are caused by mechanical ringing of acoustic resonances. \textbf{b} Time domain response to short pulses with 20 V V${_\text{pp}}$, 5 MHz repetition rate, and a 20 ns pulse width, demonstrating ultra-fast (sub-ns) tuning speed. The inset shows the normalized cross-correlation (black) between input and output signals and the auto-correlation (red) of the input signal. The right Y axis in \textbf{a}, \textbf{b} denotes the resonant wavelength shifting relative to 0 V voltage, according to a linear tuning of -0.2 pm/V. \textbf{c} 6 GHz square wave driving at the frequency where mechanical resonances disappear due to low mechanical Q. The electrical signal (gray) is measured by an oscilloscope after 20 dB attenuation of the original signal. The optical output shows clear oscillations with a frequency equal to the driving field, illustrating GHz level piezoelectric actuation. \textbf{d} Cross-section of the device after roughing and then pasting polyurethane (PU) epoxy (mixed with 3 $\upmu$m Nickle powders) on the backside of the Si substrate. \textbf{e} S$_{11}$ and \textbf{f} S$_{21}$ of the TM mode after weakening acoustic resonances. The VNA responses become smoother with only wide range envelope variation. Enabled only by AlN fundamental and second harmonic resonances (green shaded regions), broadband modulation can be achieved with 3 dB bandwidth of 250 MHz for each.}
\label{Fig:figure4}
\end{figure}

Besides the resonant microwave frequency acousto-optic modulation, the proposed stress-optical platform is capable of ultra-fast tuning of optical resonances with ultra-low power consumption (see Supplementary Material for detailed results). With the same device, bi-directional tuning can be achieved with linear tunability of -0.2 pm/V and power efficiency of 5 nW/pm. Here, the tuning speed is fully explored, showing sub-ns response time. To do this, the time domain dynamic response is recorded while applying a modulated signal. A small-signal (V$_{\text{pp}}$ = 2 V) square wave (1 MHz) is first applied in Fig. \ref{Fig:figure4}a, and the transmitted light intensity modulation is measured. Although with only 0.4 pm relative resonant wavelength shifting, the output signal shows clear separation between two switching states which is larger than the noise level. We also observe sharp peaks at tuning edges when rapidly switching from one state to the other, which is caused by the excitation of acoustic resonances. Periodic (5 MHz) short pulses with 20 ns pulse widths are also applied as seen in Fig. \ref{Fig:figure4}b. To quantitatively show the similarity between input and output signals, the normalized cross-correlation between them is calculated as shown in the inset of \ref{Fig:figure4}b. Indeed, the similarity between the auto-correlation (red) of the input signal itself and the cross-correlation (black) demonstrates ultra-fast actuation beyond the nanosecond. 

Despite the high actuation speed, the existence of HBAR modes prevents digital modulation at repetition rates beyond 1 GHz, where a flat and broadband response is usually required as in traditional optical communication. In this sense, pre-conditioning of the input signal or data post-processing is necessary to eliminate the distortion from mechanical resonances. However, due to the constant $f\cdot Q$ product in a general mechanical resonant systems, the acoustic resonances at high frequency gradually disappear due to low mechanical Q. If we apply an ultra-high repetition rate (f$_{\text{rep}}$) signal in the mechanical resonance damping-out region, we can still retrieve a clear output, getting rid of resonance induced signal distortion. This is demonstrated in Fig. \ref{Fig:figure4}c, where a 6 GHz square wave (gray curve) is applied using a programmable pattern generator and amplified by an optical modulator driver to V${_\text{pp}}$ of 7 V. The output optical modulation (orange) shows distinguishable oscillation at the same repetition rate, which suggests that the piezoelectric actuator itself can perform ultra-fast actuation, despite the fact that broadband modulation is mostly limited by acoustic resonances. Because of the lower responsivity to higher harmonic Fourier components (e.g., 12 GHz, 24 GHz) of the input square wave, the output behaves more like a 6 GHz sinusoidal wave. 

Note that, even at ultra-high f$_{\text{rep}}$, the pseudo-random Non-Return-to-Zero (NRZ) signal also contains broad frequency components besides harmonics of f$_{\text{rep}}$, which still suffer from acoustic resonances. For applications where a broad bandwidth is required, these acoustic modes should be effectively damped or even eliminated. To do this, the bottom surface of the Si substrate is first roughed by XeF$_2$ isotropic etching to diffract acoustic waves, and then polyurethane (PU) epoxy (mixed with 3 $\upmu$m Nickle powder) is pasted onto the backside for damping acoustic vibrations \cite{pinrod2018}, as shown in Fig. \ref{Fig:figure4}d. After post-processing the fabricated device, its electromechnical and optomechanical performances are recorded in Fig. \ref{Fig:figure4}e and f, respectively. To increase the signal to noise ratio of S$_{21}$ at high frequencies, the RF signal from the VNA is amplified before being applied to the actuator. From the S$_{11}$ response, one can see that nearly all resonances from the Si substrate cavity are diminished. However, the smoothly varying envelope from resonances in SiO$_2$ and AlN cavities still exists, since they remain unaffected by the post-processes. The S$_{21}$ measurement also demonstrates smoothing of the modulation spectrum, but with broad range variations. Some exceptionally small resonances are still visible below 2.5 GHz in S$_{21}$, because the acoustic wavelength at low frequencies is comparable or larger than the Si roughing scale ($\sim$5 $\upmu$m). It's worth noting here that the optical measurement is more sensitive than its electrical counterpart due to high optical Q and its signal to noise ratio. The fast roll-off of S$_{21}$ starting around 9.5 GHz is mainly limited by the optical quality factor of the TM mode, and the actual acousto-optic interaction may potentially extend to frequencies beyond 10 GHz. 

The fundamental resonance from the AlN thin film enhances the electromechanical conversion efficiency and thus optical modulation around 4.13 GHz, and thanks to its low mechanical Q, we observe broadband modulation with a 3 dB bandwidth of 250 MHz, as shown in Fig. \ref{Fig:figure4}f. The second harmonic resonance of AlN cavity is also found in S$_{21}$ around 8.7 GHz, here with a 260 MHz bandwidth. These broad bands of modulation can potentially be used to connect supperconducting circuits with optical interfaces for low-loss quantum information communication \cite{kurizki2015,midolo2018nano}. Of note, the positions of these bands can be engineered by modifying the AlN and SiO$_2$ film thickness for specific applications. For future works, dedicated design and patterning of the bottom surface can be done for better suppression of acoustic resonances from Si cavity. To further increase modulation bandwidth, more effort is needed to eliminate resonances from SiO$_2$ and AlN layers, which can be realized by matching mechanical impedance at interfaces, such as using an acoustic anti-reflection layer \cite{zhu2017perfect}, or through phononic crystal band gap engineering to suppress undesired acoustic modes \cite{lu2009phononic}. 

In conclusion, we demonstrate for the first time the integration of HBAR resonators within a nano-photonic platform. We realize a hybrid multi-functional stress-optical platform based on bulk acoustic standing and traveling waves. In the device, acoustic resonances are effectively generated by a macroscopic AlN actuator, and are used for microwave frequency acousto-optic modulation. These acoustic modes are evenly distributed over a broad frequency range up to 6 GHz. To put this work into perspective, it is worth pointing out several key differences between our platform and conventional SAW-based AOMs \cite{tadesse2014}. Firstly, in our scheme, optical structures can be buried deep inside the cladding material, which preserves the high optical Q. Secondly, the critical size of the actuator does not need to be scaled down for exciting high frequency (sub-micron acoustic wavelength) acoustic waves, since acoustic resonances primarily rely on the dimensions of the vertical stack. This enables optical lithography, high power handling, and a small footprint (due to the overlay of mechanical and optical components). Finally, the confinement of acoustic modes vertically inside the substrate guarantees low cross-talk between adjacent devices, and thus compact integration. Altogether, these advantages serve to produce a device which is not only novel, but has potential for widespread applications across a variety of fields.


\section*{Methods}
\label{Methods} 
\label{sec_method}
\subsection{Device fabrication}
The Si$_3$N$_4$ waveguides are fabricated using a subtractive process \cite{Luke:13}. The fabrication process of AlN piezoelectric actuators is described here in detail. 100 nm Mo and 1 $\mu$m polycrystalline AlN films are sputtered on SiO$_2$ cladding through foundry services (OEM Group). The AlN disk is patterned by AZ1518 photoresist, and dry etched using Cl$_2$ and BCl$_3$ in a Panasonic E620 Etcher. The dry etching of the bottom electrode (Mo) is performed using Cl$_2$ in the same Panasonic E620 Etcher. Finally, the top 100 nm of Al is evaporated by a PVD E-beam evaporator, and patterned using a standard lift-off process. The whole wafer is diced into individual chips by deep reactive ion etching (RIE) followed by chemical mechanical polishing (CMP) of the Si substrate. This enables smooth chip facets for efficient coupling of light from lensed fiber to inverse waveguide taper at the edge. This three-mask, photolithographic-only fabrication leads to a low cost and high fabrication tolerance. 

\subsection{Damping bulk acoustic resonances}
To damp and eliminate the intrinsic bulk acoustic wave resonances, the backside surface of the Si substrate is first isotropically etched by XeF$_2$ in a Xactix Xenon Difluroide E1 system, with the top surface protected by the photoresist AZ1518. The roughing will diffract acoustic waves to random directions and thus weaken constructive interference. Next, as suggested by previous work \cite{pinrod2018}, a layer of polyurethane (PU) epoxy mixed with 3 $\mu$m nickle powder is pasted at the bottom, which can damp acoustic vibration at the boundaries and absorb acoustic energy. 

\subsection{Experimental setup}
The electromechanical S$_{11}$ reflection response is measured by port 1 of a network analyzer (Agilent E8364B), where the electrical signal is applied to the device through an RF GS probe (GGB 40A-GS-150). For the optomechanical S21 measurment, around 100 $\upmu$W continuous wave (CW) light from a diode laser (Velocity Tunable Laser 6328) is edge coupled to the device using a lensed fiber via inverse taper. A -5 dBm RF electrical signal is applied from port 1 of the network analyzer to driving the piezoelectric actuator, and the light intensity modulation is detected by a 12 GHz photodiode (New Focus 1544), which is sent back to port 2 of network analyzer. 

In the high repetition rate time domain modulation measurement, a 6 GHz square wave is supplied by a programmable pattern generator (Tektronix PPG1251), which is amplified by an optical modulator driver (WJ communication SA1137-2) to 7 V V$_{\text{pp}}$ before applying to the device. The intensity modulation is detected by the same photodiode (New Focus 1544) and then recorded by a broad bandwidth oscilloscope (Tektronix DSA8200).

\begin{addendum}

\item[Acknowledgements]
This work was supported by Swiss National Science Foundation under grant agreement No. 176563 (BRIDGE), and by the Defense Advanced Research Projects Agency (DARPA), Microsystems Technology Office (MTO) under contract No. HR0011-15-C-0055 (DODOS), and by funding from the European Union’s H2020 research and innovation programme under FET Proactive grant agreement No. 732894 (HOT).
The samples were fabricated in the EPFL center of MicroNanoTechnology (CMi), and Birck Nanotechnology Center at Purdue University. The authors would like to thank Dr. Daniel E. Leaird for helping testing high frequency time domain modulation, and Mohammad Bereyhi and Dr. Ben Yu for valuable discussion.

\item[Additional information]
See Supplementary Material for supporting content. The data that support the plots are available from the corresponding authors upon reasonable request.

\item[Competing interests]
The authors declare no competing interests.

\end{addendum}
\clearpage

\title{\textbf{Supplementary Material: Hybrid Integrated Photonics Using Bulk Acoustic Resonators}}  

\author{\noindent Hao Tian$^{1}$, Junqiu Liu$^{2}$, Bin Dong$^{1}$, J Connor Skehan$^{2}$, Michael Zervas$^{2}$, Tobias J. Kippenberg$^{2,\dagger}$, Sunil A. Bhave$^{1,\dagger}$}

\begin{affiliations}
\item OxideMEMS Lab, Purdue University, 47907 West Lafayette, IN, USA
\item Institute of Physics, Swiss Federal Institute of Technology Lausanne (EPFL), 1015 Lausanne, Switzerland
\maketitle

\normalsize{$^\dagger$ E-mail: tobias.kippenberg@epfl.ch, bhave@purdue.edu}
\end{affiliations}

\section{Analytic analysis of electro-opto-mechanical response}
\subsection{Electromechanical model of HBAR mode}
\begin{figure}[t]
\centering\includegraphics[width=\textwidth]{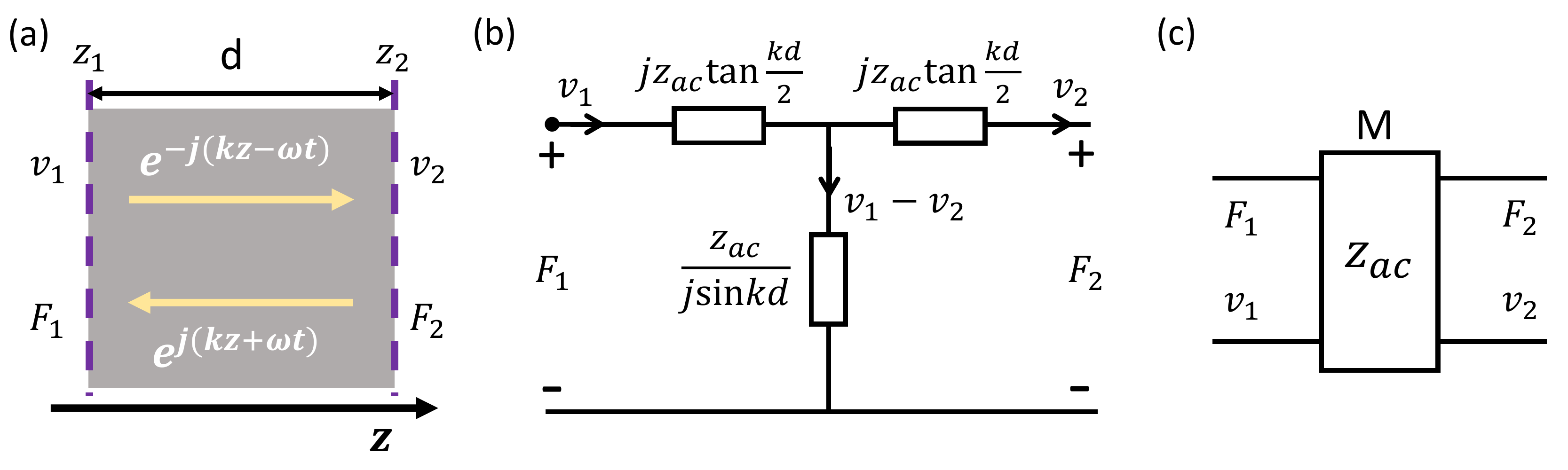}
\caption{Model of acoustic wave propagation in a non-piezoelectric material. (a) Acoustic wave propagates in forward and backward directions in a layer of non-piezoelectric material. The layer thickness is $d$ with two boundaries located at $z_1$ and $z_2$. The wave distribution is solely determined by the boundary conditions of force $F$ and velocity $v$. (b) Equivalent circuit model describing acoustic wave transmission. (c) The two boundaries can be correlated by a transfer matrix M which is a function of the acoustic impedance and the propagation length. }
\label{Fig1}
\end{figure}

To better explain the electromechanical S11 response and mechanical dispersion shown in the main text, a one-dimensional (1D) analytic electromechanical model is established by combining the well-known Mason model \cite{tirado2010bulk, zhang2003resonant} and the transfer matrix method \cite{chen2006characterization}. As shown in Fig. \ref{Fig1}(a), the acoustic wave is assumed to propagate bidirectionally along the z axis due to acoustic reflection at interfaces, such that the mechanical displacement $u(z,t)$ can be expressed as \cite{tirado2010bulk}:
\begin{equation}
    u(z,t)=A^+e^{-j(kz-\omega t)}+A^{-}e^{j(kz+\omega t)}
\end{equation}
where, $\omega$ and $k=\omega /v_{ac}$ ($v_{ac}$ is acoustic velocity) are the frequency and acoustic wave number respectively, and $A^+$ and $A^-$ are amplitudes for the bidirectional propagating waves. The wave is then related to the velocities $v$ and forces $F$ (or stress $\sigma$) at the two surfaces at $z_1$ and $z_2$, working as boundary conditions \cite{tirado2010bulk}:
\begin{align}
    v_1 &=\frac{\text{d}u(z_1)}{\text{d}t}=j\omega (A^+e^{-jkz_1}+A^{-}e^{jkz_1})\\
    v_2 &=\frac{\text{d}u(z_2)}{\text{d}t}=j\omega (A^+e^{-jkz_2}+A^{-}e^{jkz_2})
\end{align}
\begin{align}
    F_1 &=Sc\frac{\text{d}u(z_1)}{\text{d}z}=-jSck(A^+e^{-jkz_1}-A^{-}e^{jkz_1})\\
    F_2 &=Sc\frac{\text{d}u(z_2)}{\text{d}z}=-jSck(A^+e^{-jkz_2}-A^{-}e^{jkz_2})
\end{align}

S is the surface area, and c is the stiffness coefficient of the material . The prefactor related with time is omitted for simplicity. After some brief algebra, the forces can be expressed as the combination of velocities as:
\begin{align}
    F_1 &=\frac{Z_{ac}}{j\text{sin}(kd)}(v_1-v_2)+jZ_{ac}\text{tan}(\frac{kd}{2})v_1\\
    F_2 &=\frac{Z_{ac}}{j\text{sin}(kd)}(v_1-v_2)-jZ_{ac}\text{tan}(\frac{kd}{2})v_2
\end{align}
where $Z_{ac}$ (=$S\rho v_{ac}$, where $\rho$ is the material density) is the acoustic impedance of the material, and $d$ is the thickness. Interestingly, if we treat force and velocity as voltage and current, an equivalent circuit model can be built which satisfies Eq. (6-7) according to Kirchhoff's law, as shown in Fig. \ref{Fig1}(b). The circuit consists of three resistors with impedance as labeled in Fig. \ref{Fig1}(b). Since the force and velocity must be continuous at the boundary between two different layers, the circuit model makes it easy to cascade different layers by connecting their corresponding circuits. 

From Eq. (6-7), two adjecent boundaries can be related using a transfer matrix M as \cite{chen2006characterization}:
\begin{equation}
    \begin{bmatrix}
    F_1\\v_1
    \end{bmatrix}=\begin{bmatrix}
    \text{cos}(kd) & jZ_{ac}\text{sin}(kd)\\j\text{sin}(kd)/Z_{ac} & \text{cos}(kd)
    \end{bmatrix}
    \begin{bmatrix}
    F_2\\v_2
    \end{bmatrix}=M\begin{bmatrix}
    F_2\\v_2
    \end{bmatrix}
\end{equation}
In this way, each layer can be represented by its characteristic transfer matrix M [Fig. \ref{Fig1}(c)], and the relation between any two boundaries can be connected by multiplying the transfer matrix of each layer in between. The boundary condition at each interface can thus be determined from the very end boundaries of the entire stack structure which, for a general mechanical structure, satisfy the free boundary condition where the force is zero ($F=0$), or the fixed boundary condition where the velocity (or equivalently the displacement) is zero ($v=0$). After knowing the boundary conditions, the acoustic wave distribution in each layer can be determined from Eq. (2-5) by solving for $A^+$ and $A^-$. This is known as transfer matrix method for solving one-dimensional propagation of acoustic waves in multiple layer structures, which is suitable for our vertical stack structure of HBAR mode. 

After the derivation of acoustic wave propagation, we are now ready for the model of acoustic wave excitation through a piezoelectric actuator. As we apply voltage to the piezoelectric material, the electric field will generate stress inside the film, which in turn builds up extra charges at the surfaces and change the electric field accordingly. The interplay between stress and electric field can be related as \cite{tirado2010bulk}:
\begin{align}
    \sigma &=c^E\varepsilon+eE\\
    D &=e\varepsilon+\epsilon E
\end{align}
where $c^E$ is the stiffness coefficient under constant E, $e$ is piezoelectric coefficient, $\epsilon$ is dielectric constant, $\varepsilon$ is strain, E is electric field, and D is electric displacement. In general, the coefficients are matrices which correlate the mechanical and electric field in different directions. In our case, we consider only the terms related to the z direction. 

\begin{figure}[htbp]
\centering\includegraphics[width=12.5 cm]{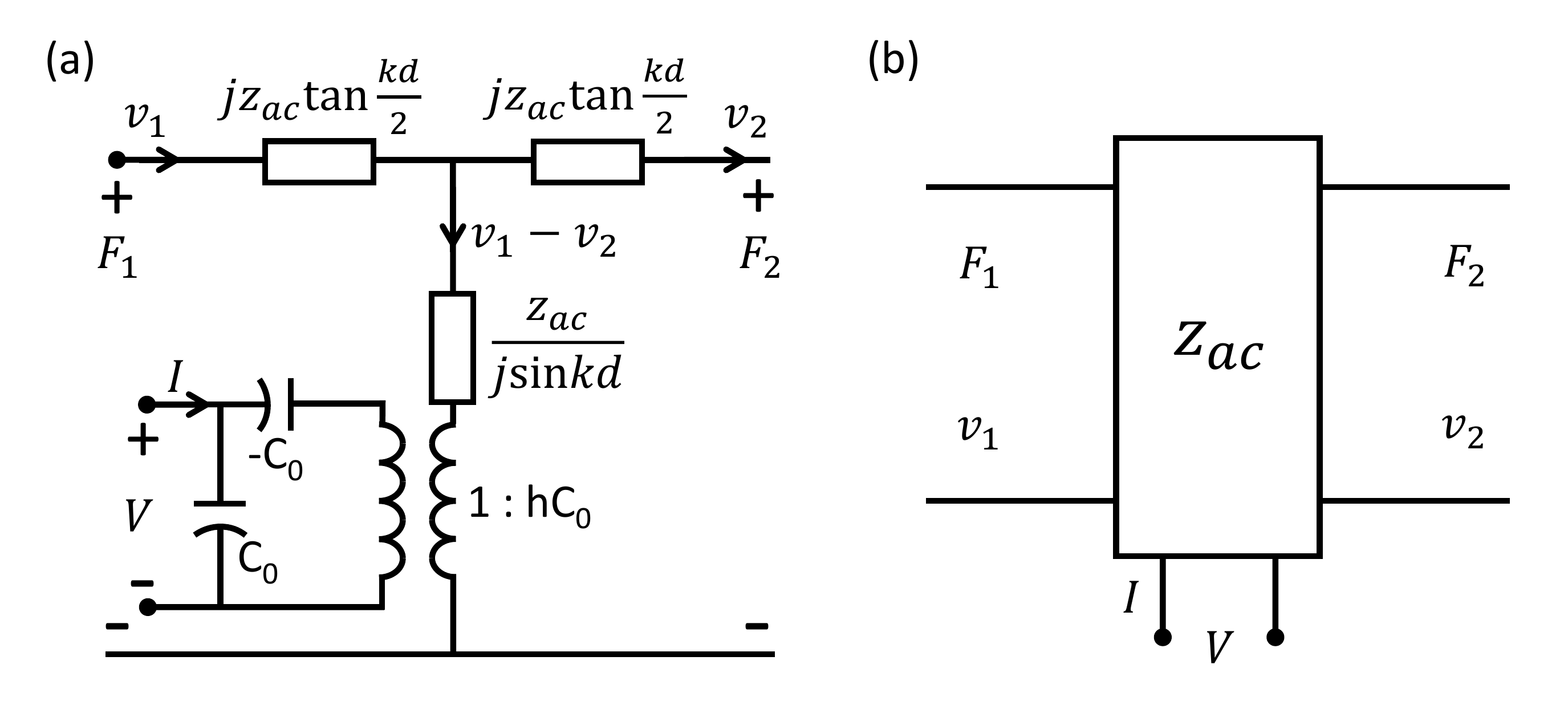}
\caption{Model of acoustic wave propagation in a piezoelectric material. (a) Equivalent circuit Mason model that describes the excitation and propagation of acoustic waves in the piezoelectric layer. The three resistors represent the propagation of acoustic waves, while the transformer represents energy conversion between electrical and mechanical domain. (c) Three ports representation of the piezoelectric actuator. }
\label{Fig2}
\end{figure}

By performing a similar procedure as before, the velocities and forces generated at the boundaries can be calculated from the applied external voltage and current through \cite{tirado2010bulk}:
\begin{align}
    F_1 &=\frac{Z_{ac}}{j\text{sin}(kd)}(v_1-v_2)+jZ_{ac}\text{tan}(\frac{kd}{2})v_1+\frac{h}{j\omega}I\\
    F_2 &=\frac{Z_{ac}}{j\text{sin}(kd)}(v_1-v_2)-jZ_{ac}\text{tan}(\frac{kd}{2})v_2+\frac{h}{j\omega}I\\
    V &=\frac{1}{j\omega C_0}[I+hC_0(v_1-v_2)]
\end{align}
where $C_0$ (= $\epsilon S/d$) is the intrinsic capacitance of the piezoelectric actuator, and $h=e/\epsilon$ is a constant related with the material properties. Based on these equations, an equivalent circuit model can be established as shown in Fig. \ref{Fig2}(a), which is the so-called Mason model \cite{tirado2010bulk}. Compared with the previous circuit in Fig. \ref{Fig1}(b) which does not consider the piezoelectric effect, a transformer is added to the middle branch with a ratio of $1:hC_0$, which then connects to the external power source through series and parallel capacitances $C_0$. The series capacitance has a negative sign which indicates that its current will combine with the external current $I$ and go through the parallel capacitance. This is to be consistent with Eq. (13). The other resistors describe the acoustic wave propagation in the piezoelectric layer as before. The piezoelectric layer can be treated as a three port component as shown in Fig. \ref{Fig2}(b), where the mechanical ports are dependent on $I$-$V$ port. In transfer matrix method, this active component introduces additional boundary conditions through the electric port. 

\begin{figure}[htbp]
\centering\includegraphics[width=\textwidth]{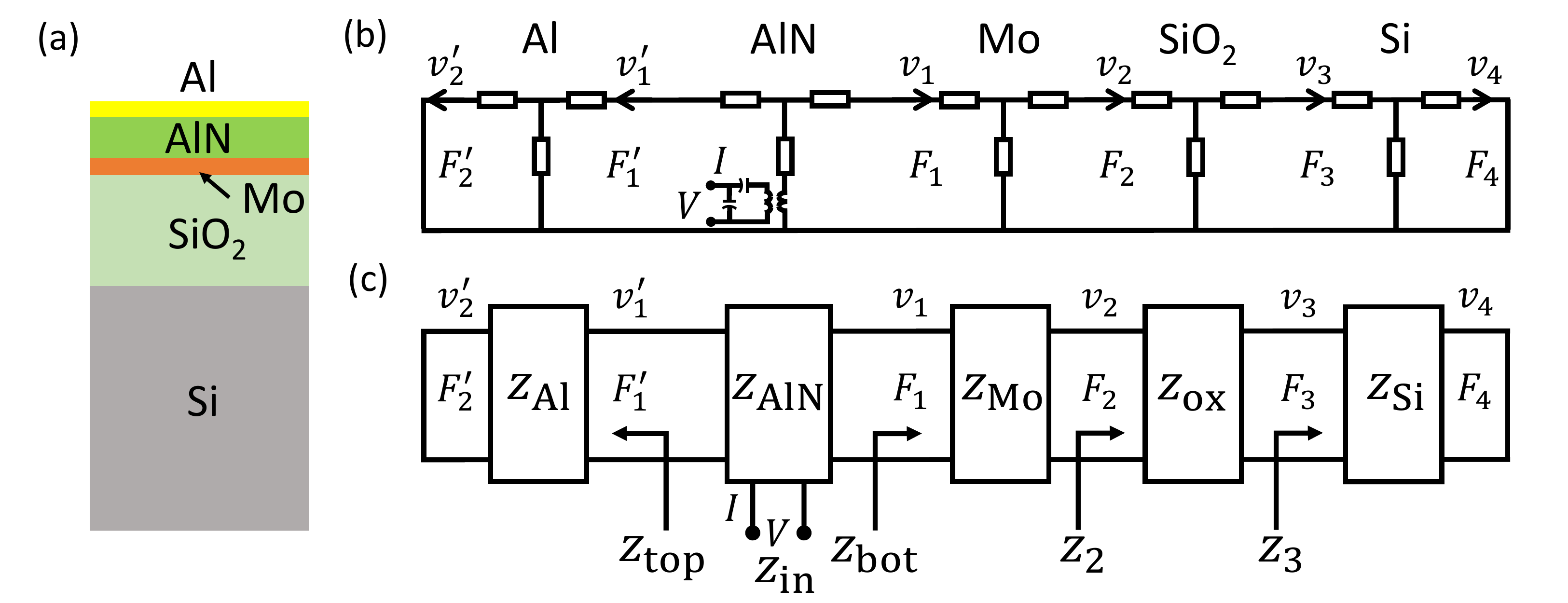}
\caption{Electromechanical model of the actual device in this work. (a) Vertical stacking structure of the whole device. (b) Equivalent circuit model by connecting adjacent layers. The end ports are shorted as required by the free boundary condition. (c) Transfer matrix chain that connects each interface. The input impedance at each port can be correlated and calculated by the multiplication of matrices in between. }
\label{Fig3}
\end{figure}

As we have the circuit model and transfer matrix of each layer, the actual device as described in the main text and shown in Fig. \ref{Fig3}(a) can be modeled by simply connecting each adjacent layer. Fig. \ref{Fig3}(b) shows the equivalent circuit of the whole device. Free boundary conditions at the top and bottom surfaces are employed such that the forces $F'_2$ and $F_4$ are zero, which correspond to an electric short in the circuit. The impedance looking into one interface can be defined as $Z=F/v$. By utilizing Eq. (8), and assuming $F'_2=0$ and $F_4=0$, the impedance at port 1$'$ and 3 from the top Al electrode and bottom Si substrate can be calculated easily as:
\begin{align}
    Z_{\text{top}} &=jZ_{\text{Al}}\text{tan}(k_{\text{Al}}d_{\text{Al}})\\
    Z_{3} &=jZ_{\text{Si}}\text{tan}(k_{\text{Si}}d_{\text{Si}})
\end{align}
where $Z_{\text{Al}}$ and $Z_{\text{Si}}$ are the acoustic impedance of Al and Si respectively. Similarly, by multiplying matrices of cascaded layers, the impedance at ports 2 and 1 can be calculated:
\begin{align}
    Z_{2} &=j\frac{Z_{\text{Si}}\text{tan}(k_{\text{Si}}d_{\text{Si}})+Z_{\text{ox}}\text{tan}(k_{\text{ox}}d_{\text{ox}})}{1-(Z_{\text{Si}}/Z_{\text{ox}})\text{tan}(k_{\text{Si}}d_{\text{Si}})\text{tan}(k_{\text{ox}}d_{\text{ox}})}\\
    Z_{\text{bot}} &=\frac{Z_2+jZ_{\text{Mo}}\text{tan}(k_{\text{Mo}}d_{\text{Mo}})}{1+j(Z_2/Z_{\text{Mo}})\text{tan}(k_{\text{Mo}}d_{\text{Mo}})}
\end{align}
where $Z_{\text{ox}}$ and $Z_{\text{Mo}}$ are the acoustic impedance of SiO$_2$ and Mo respectively. From Eq. (11-13), the electrical impedance $Z_{\text{in}}$ ($=V/I$) can be obtained by \cite{tirado2010bulk, zhang2003resonant}:
\begin{equation}
    Z_{\text{in}}=\frac{1}{j\omega C_0}\left\{1-\frac{k_{\text{t}}^2}{k_{\text{AlN}}d_{\text{AlN}}}\frac{(z_{\text{top}}+z_{\text{bot}})\text{sin}(k_{\text{AlN}}d_{\text{AlN}})+j2[1-\text{cos}(k_{\text{AlN}}d_{\text{AlN}})]}{(z_{\text{top}}+z_{\text{bot}})\text{cos}(k_{\text{AlN}}d_{\text{AlN}})+j(1+z_{\text{top}}z_{\text{bot}})\text{sin}(k_{\text{AlN}}d_{\text{AlN}})}\right\}
\end{equation}
where, $k_{\text{t}}^2$ is the intrinsic electromechanical coupling coefficient of AlN which is 6.5\%, $z_{\text{top}}$ ($=Z_{\text{top}}/Z_{\text{AlN}}$) and $z_{\text{bot}}$ ($=Z_{\text{bot}}/Z_{\text{AlN}}$) are the impedance from the top and bottom side of AlN which are normalized by the AlN acoustic impedance $Z_{\text{AlN}}$. 

\subsection{Electromechanical S$_{11}$ reflection parameter}
\begin{figure}[htbp]
\centering\includegraphics[width=11 cm]{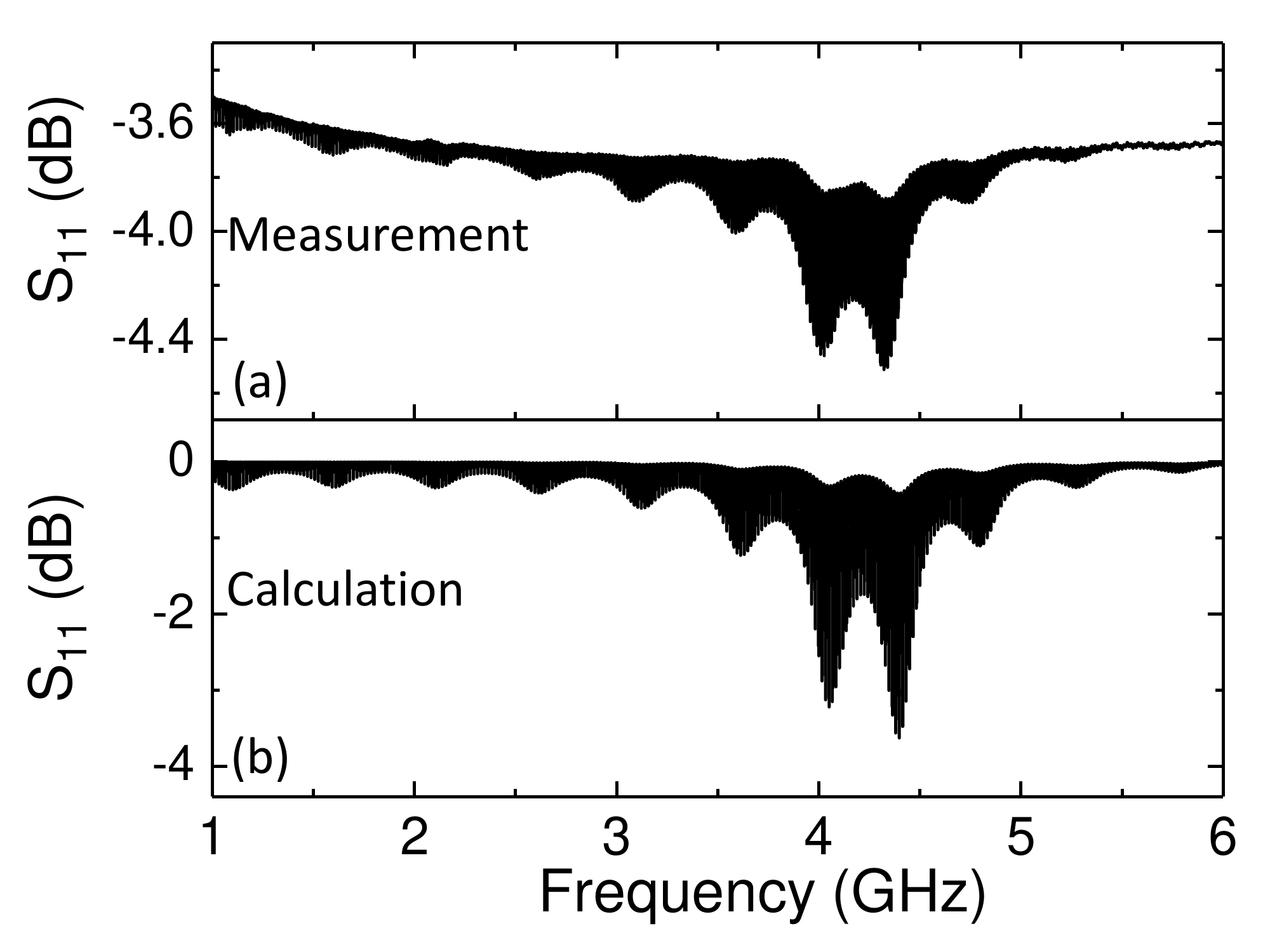}
\caption{Measured (a) and calculated (b) S$_{11}$ reflection parameter, showing much similarity in terms of varied envelope and its period.}
\label{Fig4}
\end{figure}

By applying actual material properties as summarized in Table \ref{table1}, the electrical input impedance can be calculated, from which the S$_{11}$ response can be calculated as:
\begin{equation}
    S_{11}=\frac{Z_0-Z_{\text{in}}}{Z_0+Z_{\text{in}}}
\end{equation}
where $Z_0$ (50 $\Omega$) is the standard normalized impedance of the network analyzer. The results are shown in Fig. \ref{Fig4} which demonstrates similarity between measurements and calculations from the electromechanical model. The difference in magnitude is mainly introduced by the calibration and parasitic capacitance from the probe landing during the electrical experiments. The variation of envelope caused by the coupling between Si, SiO$_2$ and AlN cavities is well captured by the model. As mentioned in the main text, the node of the envelope corresponds to a SiO$_2$ resonance. This is because, at the SiO$_2$ resonance, more acoustic energy is confined in SiO$_2$ which is softer and has smaller acoustic impedance as compared to Si. 

\begin{table}
    \centering
    \caption{Material properties of each layer employed in the analytic model}
    \begin{tabular}{ p{3cm}|p{3.5cm}|p{3cm}|p{3.5cm}  }
 \hline
 Material  & Density $\rho$ (kg/m$^3$) & Velocity $v$ (m/s) & Thickness $d$ ($\upmu$m)\\
 \hline
 Al   & 2700    &6300&   0.1\\
 AlN &   3300  & 11050   &0.92\\
 Mo & 10200 & 6636&  0.1\\
 SiO$_2$    &2200 & 5640&  5.44\\
 Si &   2329  & 8430 &231.5\\
 \hline
\end{tabular}
    \label{table1}
\end{table}

\begin{table}
    \centering
    \caption{Effective cavity length at different resonant conditions. The effective boundary condition at each interface is compared. }
    \begin{tabular}{ p{1.5cm}|p{1.5 cm}|p{1.7 cm}|p{1.8cm}|p{1.8 cm}|p{5.4 cm}  }
 \hline
 $k_{\text{AlN}}d_{\text{AlN}}$  & $k_{\text{ox}}d_{\text{ox}}$ & $k_{\text{Si}}d_{\text{Si}}$ & AlN--SiO$_2$  & SiO$_2$--Si & Effective length\\
 \hline
 $p\pi$   & $n\pi$    & $m\pi$ &   Free & Free & $d_{\text{Si}}+\frac{\rho _{\text{ox}}}{\rho _{\text{Si}}}d_{\text{ox}}+\frac{\rho _{\text{AlN}}}{\rho _{\text{Si}}}d_{\text{AlN}}$\\
 $p\pi$  & $n\pi +\pi /2$    & $m\pi +\pi /2$ &   Free & Fixed & $d_{\text{Si}}+\frac{Z_{\text{Si}}^2}{Z_{\text{ox}}^2}\left(\frac{\rho _{\text{ox}}}{\rho _{\text{Si}}}d_{\text{ox}}+\frac{\rho _{\text{AlN}}}{\rho _{\text{Si}}}d_{\text{AlN}}\right)$\\
 $p\pi +\pi /2$ & $n\pi +\pi /2$    & $m\pi $ &   Fixed & Free & $d_{\text{Si}}+\frac{\rho _{\text{ox}}}{\rho _{\text{Si}}}d_{\text{ox}}+\frac{Z_{\text{ox}}^2}{Z_{\text{AlN}}^2}\frac{\rho _{\text{AlN}}}{\rho _{\text{Si}}}d_{\text{AlN}}$\\
 $p\pi +\pi /2$ & $n\pi $    & $m\pi +\pi /2$ &   Fixed & Fixed & $d_{\text{Si}}+\frac{Z_{\text{Si}}^2}{Z_{\text{ox}}^2}\left(\frac{\rho _{\text{ox}}}{\rho _{\text{Si}}}d_{\text{ox}}+\frac{Z_{\text{ox}}^2}{Z_{\text{AlN}}^2}\frac{\rho _{\text{AlN}}}{\rho _{\text{Si}}}d_{\text{AlN}}\right)$\\
 \hline
\end{tabular}
    \label{table2}
\end{table}

\subsection{Mechanical dispersion analysis}
\begin{figure}[htbp]
\centering\includegraphics[width=\textwidth]{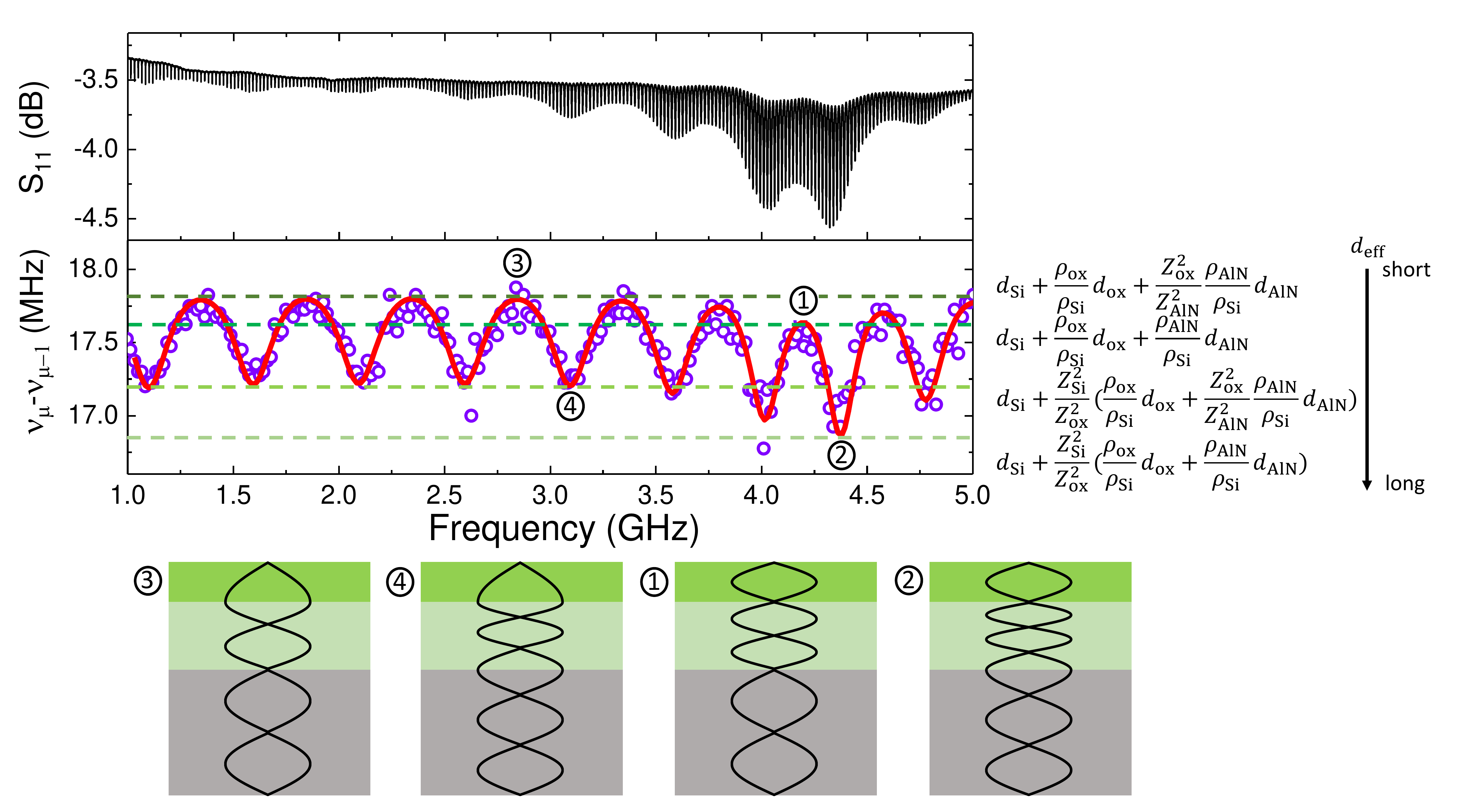}
\caption{Four different regions are labeled by number in the order from top to bottom corresponding to Table \ref{table2}. The green dashed lines denote the FSR of each region and the corresponding effective length $d_{\text{eff}}$ is labeled on the right. The bottom insets show the schematics of the acoustic stress wave distribution for each region, illustrating the locations of interfaces relative to the stress wave. }
\label{Fig5}
\end{figure}

Provided the precision of the model demonstrated in last section, we can rely on the model and analyze the mechanical dispersion by calculating resonant frequencies. By letting the denominator in Eq. (18) equal zero, we retrieve the parallel resonant frequencies corresponding to maximum resistance \cite{zhang2003resonant}:
\begin{equation}
    (z_{\text{top}}+z_{\text{bot}})\text{cos}(k_{\text{AlN}}d_{\text{AlN}})+j(1+z_{\text{top}}z_{\text{bot}})\text{sin}(k_{\text{AlN}}d_{\text{AlN}})=0
\end{equation}
The dispersion equation is transcendental and can be solved numerically. The FSR variation and higher order dispersion can thus be calculated from resonant frequencies as shown in the main text. To get a feeling of why and how the FSR varies, an analytic expression of FSR can be derived under simplified assumptions. Specifically, we can assume the metal thickness is much smaller than the acoustic wavelength, such that their impedance is nearly zero. Under this assumption, the dispersion equation Eq. (20) can be simplified to:
\begin{equation}
    Z_{\text{Si}}\text{tan}(k_{\text{Si}}d_{\text{Si}})+\frac{Z_{\text{ox}}\text{tan}(k_{\text{ox}}d_{\text{ox}})+Z_{\text{AlN}}\text{tan}(k_{\text{AlN}}d_{\text{AlN}})}{1-\frac{Z_{\text{AlN}}}{Z_{\text{ox}}}\text{tan}(k_{\text{AlN}}d_{\text{AlN}})\text{tan}(k_{\text{ox}}d_{\text{ox}})}=0
\end{equation}
The equation is arranged in the way that SiO$_2$ and AlN are combined and work together as an external cavity coupled to the Si cavity. The FSR can be divided into four regions in terms of resonance and anti-resonance of SiO$_2$ and AlN cavities (see Table \ref{table2} and Fig. \ref{Fig5}). 

Firstly, let's consider frequencies around AlN resonance, where $k_{\text{AlN}}d_{\text{AlN}}=p\pi +\delta _{\text{AlN}}$, with $p$ an integer number and $\delta _{\text{AlN}}$ denoting a small deviation. When the SiO$_2$ is at resonance such that the second term in Eq. (21) is near zero, $k_{\text{ox}}d_{\text{ox}}=n\pi +\delta _{\text{ox}}$. In this case, Si cavity must satisfy $k_{\text{Si}}d_{\text{Si}}=m\pi +\delta _{\text{Si}}$, such that Eq. (21) becomes \cite{zhang2003resonant}: 
\begin{equation}
    Z_{\text{Si}}\delta _{\text{Si}}+Z_{\text{ox}}\delta _{\text{ox}}+Z_{\text{AlN}}\delta _{\text{AlN}}=0
\end{equation}
From Eq. (22) we derive the relation between the three cavities as:
\begin{equation}
    k_{\text{Si}}d_{\text{Si}}=m\pi -\frac{Z_{\text{ox}}}{Z_{\text{Si}}}(k_{\text{ox}}d_{\text{ox}}-n\pi)-\frac{Z_{\text{AlN}}}{Z_{\text{Si}}}(k_{\text{AlN}}d_{\text{AlN}}-p\pi)
\end{equation}
where $k_{\text{Si}}=2\pi f_m/v_{\text{Si}}$, $k_{\text{ox}}=2\pi f_m/v_{\text{ox}}$, $k_{\text{AlN}}=2\pi f_m/v_{\text{AlN}}$. This is also true for the $m-1$ mode. Note that the mode order $p$ and $n$ of AlN and SiO$_2$ will not change for the $m$ and $m-1$ modes. By taking the frequency difference between the $m$ and $m-1$ modes, we can get the local FSR after some algebra as: 
\begin{equation}
    \Delta f=\Delta f_0\frac{d_{\text{Si}}}{d_{\text{Si}}+\frac{\rho _{\text{ox}}}{\rho _{\text{Si}}}d_{\text{ox}}+\frac{\rho _{\text{AlN}}}{\rho _{\text{Si}}}d_{\text{AlN}}}
\end{equation}
where $\Delta f_0=v_{\text{Si}}/(2d_{\text{Si}})$ is the original FSR of the Si cavity. This new FSR is smaller than the original FSR since the SiO$_2$ and AlN extends the effective cavity length. It is interesting to note that this extension is not simply the physical length of each layer but the effective length weighted by its density relative to Si. 

Following a similar procedure, the FSR at SiO$_2$'s anti-resonance and around AlN's resonance can be obtained. In this time, the three cavities satisfy:
\begin{align}
    k_{\text{AlN}}d_{\text{AlN}} &=p\pi +\delta _{\text{AlN}}\\
    k_{\text{ox}}d_{\text{ox}} &=n\pi +\pi /2+\delta _{\text{ox}}\\
    k_{\text{Si}}d_{\text{Si}} &=m\pi +\pi /2+\delta _{\text{Si}}
\end{align}
By inserting them into Eq. (21), it becomes \cite{zhang2003resonant}:
\begin{equation}
    -\frac{Z_{\text{Si}}}{\delta _{\text{Si}}}+\frac{-\frac{Z_{\text{ox}}}{\delta _{\text{ox}}}+Z_{\text{AlN}}\delta _{\text{AlN}}}{1+\frac{Z_{\text{AlN}}\delta _{\text{AlN}}}{Z_{\text{ox}}\delta _{\text{ox}}}}=0
\end{equation}
Since $Z_{\text{AlN}}\delta _{\text{AlN}}$ is a very small term, we can ignore it and get:
\begin{equation}
    \delta _{\text{Si}}=-\frac{Z_{\text{Si}}}{Z_{\text{ox}}}\delta _{\text{ox}}-\frac{Z_{\text{Si}}Z_{\text{AlN}}}{Z_{\text{ox}}^2}\delta _{\text{AlN}}
\end{equation}
We can then retrieve the relation between the three cavities as:
\begin{equation}
    k_{\text{Si}}d_{\text{Si}}=m\pi +\pi /2-\frac{Z_{\text{Si}}}{Z_{\text{ox}}}(k_{\text{ox}}d_{\text{ox}} -n\pi -\pi /2)-\frac{Z_{\text{Si}}Z_{\text{AlN}}}{Z_{\text{ox}}^2}(k_{\text{AlN}}d_{\text{AlN}} -p\pi)
\end{equation}
By taking the frequency difference between $f_m$ and $f_{m-1}$, we can find the local FSR as:
\begin{equation}
    \Delta f=\Delta f_0\frac{d_{\text{Si}}}{d_{\text{Si}}+\frac{Z_{\text{Si}}^2}{Z_{\text{ox}}^2}\left(\frac{\rho _{\text{ox}}}{\rho _{\text{Si}}}d_{\text{ox}}+\frac{\rho _{\text{AlN}}}{\rho _{\text{Si}}}d_{\text{AlN}}\right)}
\end{equation}

By comparing Eq. (24) and (31), it can be seen that the extra effective length due to SiO$_2$ and AlN is now multiplied by the square of the ratio between the acoustic impedance of Si and SiO$_2$. Since SiO$_2$ has a smaller impedance than Si, the effective length is longer than before, and the FSR is thus smaller . 

The cases where the AlN is near its anti-resonance condition $k_{\text{AlN}}d_{\text{AlN}}=p\pi +\pi /2+\delta _{\text{AlN}}$ can be calculated in a similar manner which are summarized in Table \ref{table2}. Note the SiO$_2$ resonance is defined as when there is no Si, and now its resonant condition becomes $k_{\text{ox}}d_{\text{ox}}=n\pi +\pi /2+\delta _{\text{ox}}$ due to the $\pi /2$ phase introduced from the AlN--SiO$_2$ interface. Similar as before, the length (FSR) is longer (smaller) around SiO$_2$'s anti-resonance as compared to its resonance. By comparing cases between AlN resonance and anti-resonance regions, there is an additional factor of the ratio square between the impedance of oxide and AlN around AlN anti-resonance. Since AlN has a higher impedance than SiO$_2$, the effective length (FSR) around AlN anti-resonance regions is shorter (larger) than the corresponding region around AlN resonance. Based on the effective length expressions, we can find that, for two coupled mechanical cavities, if the impedance of the small cavity is smaller (e.g., Si and SiO$_2$), the effective length (FSR) is shorter (larger) around the small cavity resonance than its anti-resonance. This is also true for the reverse case (e.g., SiO$_2$ and AlN), where the small cavity has larger impedance, and the effective length (FSR) is longer (smaller) around AlN resonance \cite{zhang2003resonant}. 

The boundary conditions at the interfaces in each case are also summarized in Table \ref{table2}. The AlN--SiO$_2$ interface is free at the AlN resonance and fixed at anti-resonance. This is also true for the SiO$_2$--Si interface. The four cases are numbered as 1-4 from top to bottom in Table \ref{table2} and labeled and compared with experiments as shown in Fig. \ref{Fig5}. Around the AlN resonance, the FSR oscillates between 1 and 2, while at AlN's anti-resonance it varies between 3 and 4, which is consistent with experiments. Finally, as a rule of thumb, since the envelope and FSR are both related with the matching of acoustic impedance, the node (anti-node) of the S$_{11}$ envelope corresponds to larger (smaller) local FSR. 

\subsection{Electromechanical coupling and acousto-optic overlap}
\begin{figure}[htbp]
\centering\includegraphics[width=\textwidth]{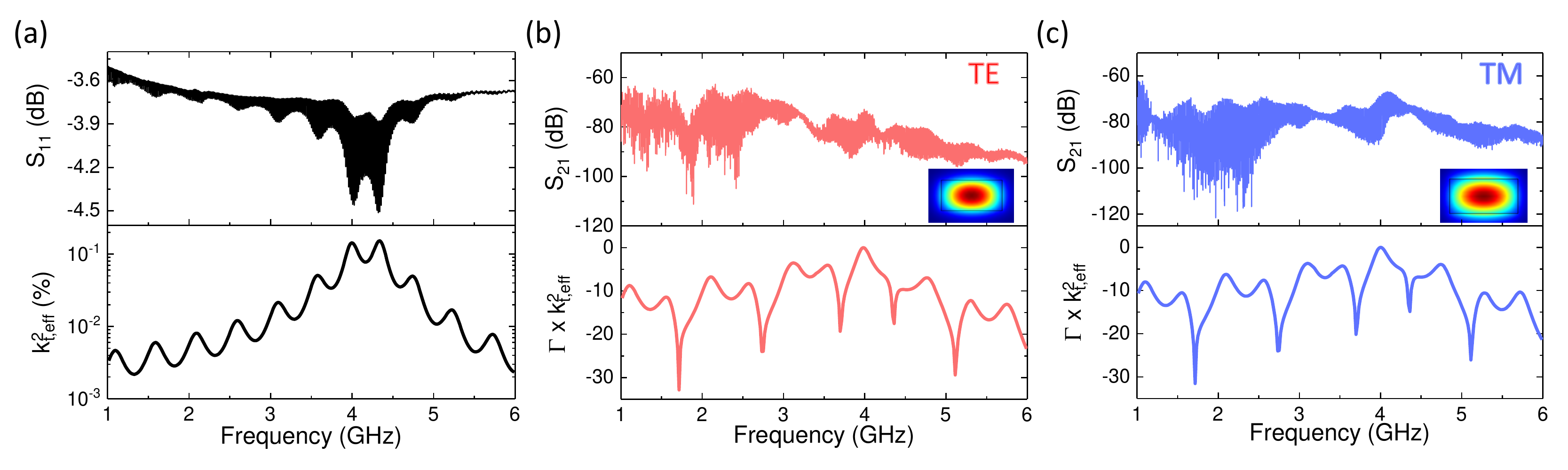}
\caption{(a) Measured S$_{11}$ response (top) and calculated effective electromechanical coupling coefficient $k_{t,eff}^2$ (bottom). The coupling reaches a maximum value of 0.15\% around 4.3 GHz. (b-c) Measured S$_{21}$ response of the TE and TM mode (top) and corresponding normalized product of acousto-optic overlap and $k_{t,eff}^2$ (bottom). When the node of the acoustic stress wave is located at the center of the waveguide, the acousto-optic overlap integral $\Gamma$ becomes zero and thus causes notches in the S$_{21}$ response. }
\label{Fig6}
\end{figure}

The model can also be used to estimate the optomechanical S$_{21}$ response which is dependent on both electromechanical coupling efficiency and acousto-optic overlap. The stress field distribution in AlN will strongly influence the effective electromechanical coupling, which can be estimated by following the method, as found in the literature \cite{zhang2003resonant}:
\begin{equation}
    k_{t,eff}^2=\frac{\pi ^2}{4}\frac{f_s}{f_p}(1-\frac{f_s}{f_p})
\end{equation}
where $f_p$ is the parallel resonant frequency when the denominator of Eq. (18) equals zero, and $f_s$ is the series resonant frequency when the numerator of Eq. (18) equals zero. $k_{t,eff}^2$ is thus calculated as shown in Fig. \ref{Fig6}(a). A maximum value of 0.15\% is reached around 4.3 GHz where the AlN resonance is located. It varies with a envelope similar as the S$_{11}$ measurement. 

In addition to elecromechanical conversion efficiency, the S21 also depends on acousto-optic overlap which determines the modulation of the optical resonant frequency. According to the perturbation theory, the relative change of resonant frequency can be related with the modulation of refractive index distribution as \cite{tadesse2014}:
\begin{equation}
    \frac{\Delta \omega}{\omega}\approx -\frac{\iint \Delta n(x,y)|\textbf{E}(x,y)|^2\text{d}x\text{d}y}{\iint n(x,y)|\textbf{E}(x,y)|^2\text{d}x\text{d}y}
\end{equation}
where the perturbation of refractive index is caused by the induced stress through the stress-optical effect, and is proportional to the stress via the stress-optical coefficient. In our specific case, the stress is dominated by the vertical stress $\sigma _z$. Therefore, the normalized acousto-optic overlap can be approximately estimated as \cite{tadesse2014}:
\begin{equation}
    \Gamma =\frac{\iint \sigma _z(z)|E(r,z)|^2\text{d}r\text{d}z}{\iint |E(r,z)|^2\text{d}r\text{d}z}
\end{equation}
where $\sigma _z$ is assumed to be dependent only on $z$ and uniform in the $r$ direction. In this way, the distribution of $\sigma _z$ can be calculated from the aforementioned 1-D acoustic model. The electric field takes only the dominant component, that is $E_r$ for TE mode and $E_z$ for TM mode. 

To better compare with the measured optomechanical S$_{21}$ response, the normalized product of $\Gamma$ and $k_{t,eff}^2$ is plotted in Fig. \ref{Fig6} (b-c) for TE and TM modes. Their product can help us to explain the variation of S$_{21}$ qualitatively. For instance, when the node of a stress wave locates at the center of the waveguide, the overlap integral approaches zero due to the vertical symmetry of the optical mode. This causes notches in the S$_{21}$ response, such as the decreasing of S$_{21}$ near 2 GHz of the TE mode. The nearly periodic variation of the envelope of S$_{21}$ is due to the modulation of $k_{t,eff}^2$. Due to the high confinement of the optical mode in the waveguide, there exhibits only a small difference of the acousto-optic overlap between the TE and TM modes. On the other hand, the difference of the measured S$_{21}$ response is mainly caused by the optical Q. Since TE mode shows nearly two times larger Q than TM mode, its response beyond 4 GHz is suppressed by entering into the resolved sideband regime. The perturbation of the local stress field due to the Si$_2$N$_4$ waveguide is not taken into account which requires 2D numerical simulation. This may lead to the difference between measurement and calculation shown in Fig. \ref{Fig6}(b-c). Despite this, the analytic model can still give us a good qualitative estimation of the electro-optomechanical response. 

\section{Investigation of different actuator shapes}

\begin{figure}[htbp]
\centering\includegraphics[width=\textwidth]{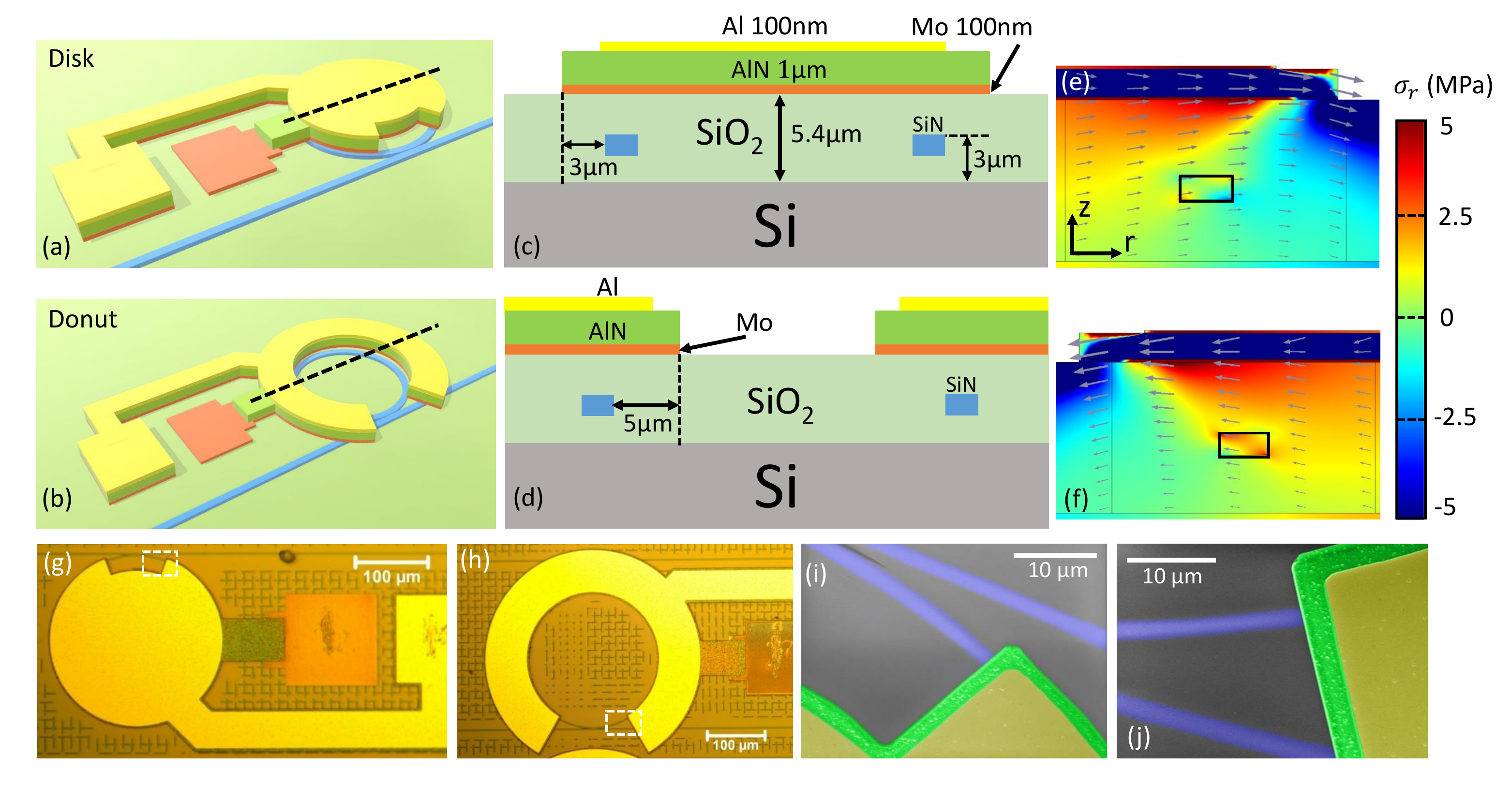}
\caption{Schematics for the designed devices with (a) Disk shape and (b) Donut shape actuators with the silicon nitride ring resonator (blue) having different relative positions. (c) and (d) are the cross-sections for the Disk and Donut devices along black dashed lines in (a) and (b), respectively. (e-f) COMSOL simulation of horizontal stress distribution around the optical waveguide under +60 V DC biasing for the Disk and Donut devices respectively. The overlaid gray arrows denote the local mechanical displacement with the biggest arrow scaled to 1 nm. (g-h) Optical microscope images of fabricated devices. (i-j) False color scanning electron micrograph (SEM) images around the actuator corner as labeled in white dashed box in (g) and (h) respectively with color similar to the cross-sections in (c-d). }
\label{Fig7}
\end{figure}

Devices with different actuator shape are fully explored, and its influence on DC optical resonance tuning and high frequency modulation is investigated. The original design with a disk shaped actuator as described in the main text is shown in Fig. \ref{Fig7}(a), which is refered as the Disk device in the following text. Since the stress mainly originates from the corner of the actuator, the microring resonator is positioned at the outer edge of the disk actuator. Under positive DC biasing on the top electrode (while the bottom electrode is always grounded), the AlN film expands and pushes the ring outwards (i.e., ring expanding) [Fig. \ref{Fig7}(e)]. Intuitively, the other design is to use a donut-shape actuator which squeezes the microring resonator if it is placed at the inner edge of the actuator, as shown in Fig. \ref{Fig7}(b). This is referred as the Donut device in the following. These two designs will show different stress distribution and static resonance tuning as demonstrated later. The cross-sections for the two designs are illustrated in Fig. \ref{Fig7}(c-d). The ring is placed 3 $\upmu$m within the edge of the Disk actuator, while 5 $\upmu$m for the Donut actuator. The final fabricated devices are shown in Fig. \ref{Fig7}(g-h) for the two designs. False color SEMs zoom-in around actuator corners in Fig. \ref{Fig7}(i-j) show the relative position between the optical microring resonator (blue) and the AlN actuator (green).

The static mechanical simulation is conducted using the finite element method (COMSOL) as shown in Fig. \ref{Fig7}(e-f), in which +60 V DC biasing is applied on the top Al layer while the bottom Mo layer is grounded. In this case, a negative electric field $E_z$ forms (points downwards) across the AlN thin film, and the AlN film will be squeezed in the z-direction and expand horizontally (positive Poisson ratio). Fig. \ref{Fig7}(e) illustrates the horizontal stress distribution and mechanical displacement around the optical waveguide at the upper right corner of Fig. \ref{Fig7}(c). As AlN expands, starting from the corner, the stress splits into two parts: extensional stress under the actuator and compressing stress at outside, and the mechanical displacement orients mainly horizontally and points outside the actuator. Similar results can be drawn for Donut device as in Fig. \ref{Fig7}(f). Depending on the position of the waveguide, it experiences extensional, compressing, or the interplay between these two. For the Disk device, the horizontal stress inside the waveguide is a mixture of extension and compression, while for the Donut device, it is mainly extension.  

\begin{figure}[htbp]
\centering\includegraphics[width=11 cm]{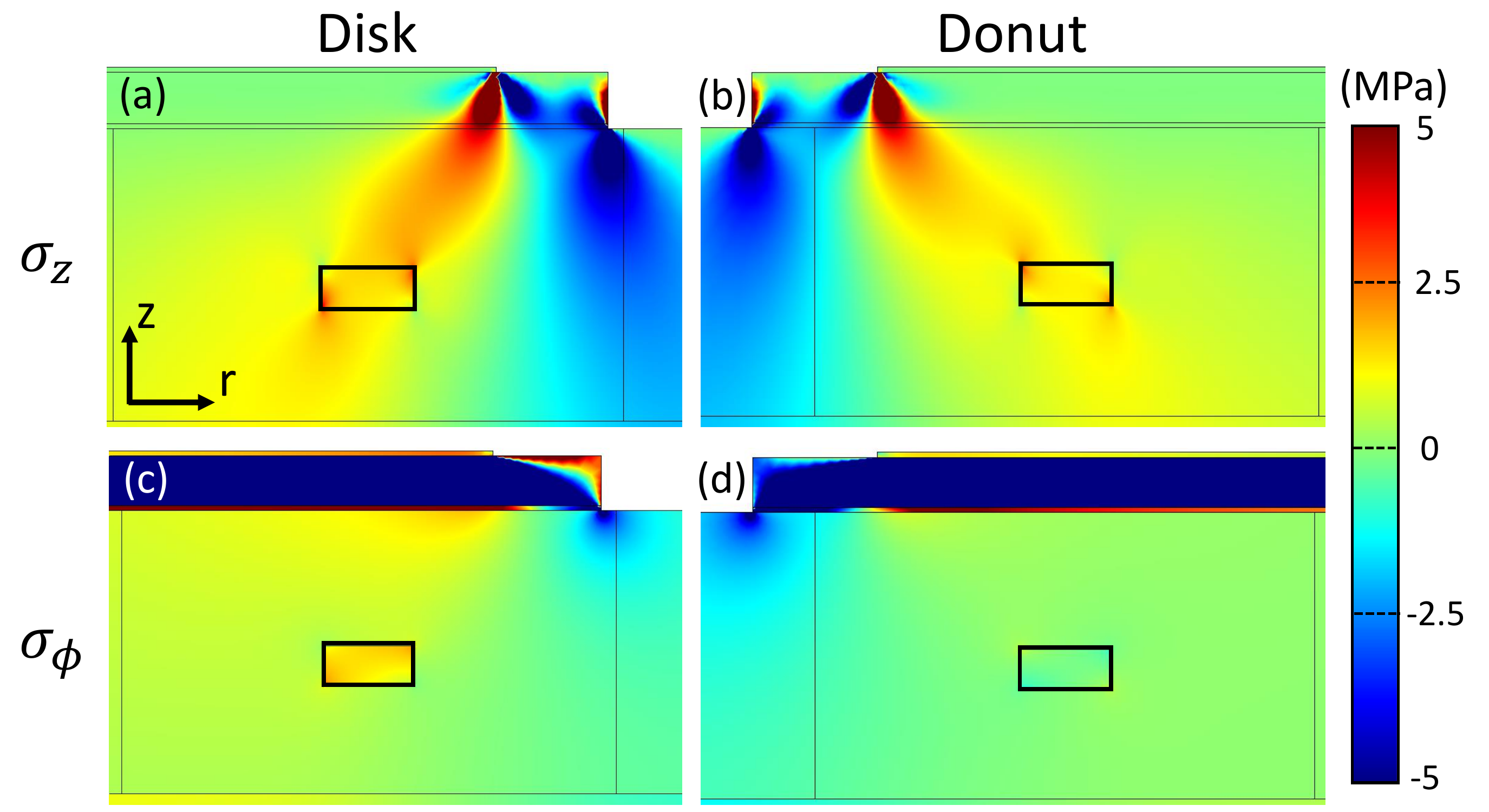}
\caption{Static stress distribution in z and $\phi$ directions. (a) and (b) show the numerical simulation of vertical (z) stress distribution under +60 V DC biasing for the Disk and Donut devices, respectively. Both of them show extensional stress around the waveguide. (c) and (d) are out of plane (tangential to optical ring) stress distributions for Disk and Donut devices. They present different sign inside the waveguide, since the optical ring of the Disk device expands whereas the Donut device squeezes. }
\label{Fig8}
\end{figure}

Besides the dominant horizontal stress, stresses in other directions will also affect the modulation on the refractive index. As shown in Fig. \ref{Fig8}(a) and (b), originating from the corner of top metal, $\sigma _z$ exhibit two main lobes with different sign. Inside the actuator, the waveguides in both cases experience extensional stress around 2 MPa. Under positive biasing, the ring of the Disk device will be pushed outwards as the actuator expands, which builds up extensional stress in the waveguide as in Fig. \ref{Fig8}(c). On the other hand, at the inner edge of the Donut actuator, the optical ring is squeezed, generating compressing stress [Fig. \ref{Fig8}(d)]. Since shear stress is found to play a less role compared with normal stress\cite{huang2003}, it is not taken into account in this study. Additionally, when applying negative voltages, the AlN actuator changes from expanding (horizontally) to shrinking, so that all the stresses in the above analysis change sign. In this way, bi-directional tuning can be achieved by reversing the applied voltage's sign, as demonstrated in the following section.

The presence of stress will change the refractive index of optical material, and affect differently for distinct polarization of the electric field of light waves, causing so-called birefringence. Numerically, they are correlated by stress-optical coefficients by \cite{huang2003}: 
\begin{align}
    n_r & = n_0-C_1\sigma _r-C_2(\sigma _{\phi}+\sigma _z) \\
    n_{\phi} & = n_0-C_1\sigma _{\phi}-C_2(\sigma _z+\sigma _r) \\
    n_z & = n_0-C_1\sigma _z-C_2(\sigma _r+\sigma _{\phi})
\end{align}
where, $n_0$ is original refractive index of the material, C$_1$ relates the refractive index and stress that are in the same direction while C$_2$ relates the two that are orthogonal. These equations are applicable to isotropic material such as amorphous Si$_3$N$_4$ from low-pressure chemical vapor deposition (LPCVD) used in this work. The lack of the stress-optical coefficient for Si$_3$N$_4$ in the literature makes it difficult to predict precisely the response of the optical ring resonator. However, it would be possible to extract the coefficients by comparing experimental tuning of optical ring resonator with simulation, which is under investigation.

\subsection{DC optical resonance tuning}

\begin{figure}[htbp]
\centering\includegraphics[width=12.5 cm]{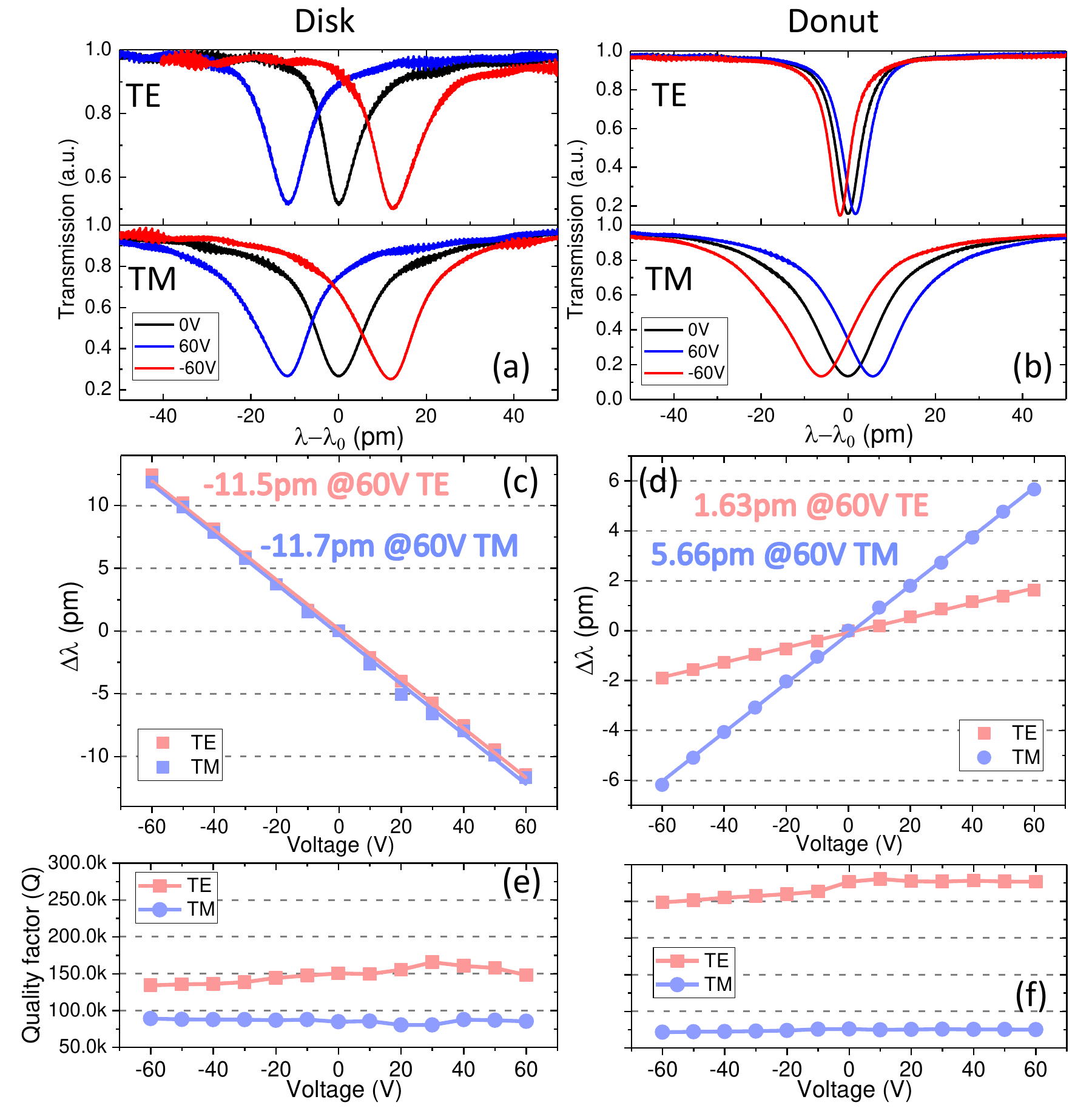}
\caption{Transmission spectrum of one resonance of TE and TM polarization mode for (a) Disk and (b) Donut device under +60 V (blue), 0 V (black), and -60 V (red). The x axis represents wavelength shifts relative to resonant wavelength $\lambda _0$ ($\sim$1550 nm) of each mode. The resonant wavelength decreases for the Disk, while it increases for the Donut under +60 V. The tuning direction reverses for opposite voltages, demonstrating bi-directional tuning. (c-d) Dependence of resonant wavelength detune on voltages for the Disk and Donut devices respectively. Experimental results (squares) show high linearity for both TE (pink) and TM (cyan) modes, with R$^{2} > 99\%$ linear fitting (straight lines). The two actuator designs show different tuning directions and tuning ranges. (e-f) Influence of piezoelectric actuation on optical quality factor Q for TE and TM modes of both designs, which verifies that the actuation will not influence the optical Q. }
\label{Fig9}
\end{figure}

Working as a Si$_3$N$_4$ ring resonator tuner, the static optical resonance tuning is performed by applying DC biasing. Fig. \ref{Fig9}(a) shows the transmission spectrum of one resonance under different voltages for both the TE and TM modes of the Disk device. One can see that as we apply a positive 60 V the resonance shifts to shorter wavelength (blue curve) relative to the original resonance (black curve), and the tuning changes direction after reversing the voltage (red curve), demonstrating bi-directional tuning ability. Also, it can be observed that the resonance dip only shifts horizontally with little changes of vertical depth, since the waveguide coupling region is unaffected by the opening section of the actuator. The dependence of resonant wavelength on voltages is summarized in Fig. \ref{Fig9}(c), showing high linearity. Both TE and TM modes demonstrate similar tuning performances with nearly -12 pm under positive 60 V. This tuning range is on a similar order or larger than the linewidth of the high optical Q resonances, which is applicable for Si$_3$N$_4$ microcomb applications such as Kerr comb generation and stabilization. 

More interestingly, due to different relative positions of ring resonator, the Donut device shows opposite behaviour: the resonant wavelength increases for positive voltages and vice versa, as can be seen in Fig. \ref{Fig9}(b) and (d). Here, the slope of tuning with respect to voltage changes from negative to positive. On the other hand, it shows much smaller tuning range with 5.66 pm for the TM mode and 1.63 pm for the TE mode under +60 V. The different tuning range of two orthogonal polarization modes, TE and TM, demonstrates tunable birefringence in an otherwise isotropic material by controlling stress, which can be utilized for polarization control \cite{xu2011polarization} or tuning the mode spacing and coupling between a pair of TE and TM modes in a ring resonator. The effect of mechanical actuation on optical Q is plotted in Fig. \ref{Fig9}(e-f), which is found to be insignificant. The TE mode shows much higher Q than the TM mode, since the TM mode extends further in the vertical direction which is prone to the absorption of the bottom metal.

\begin{figure}[t]
\centering\includegraphics[width=\textwidth]{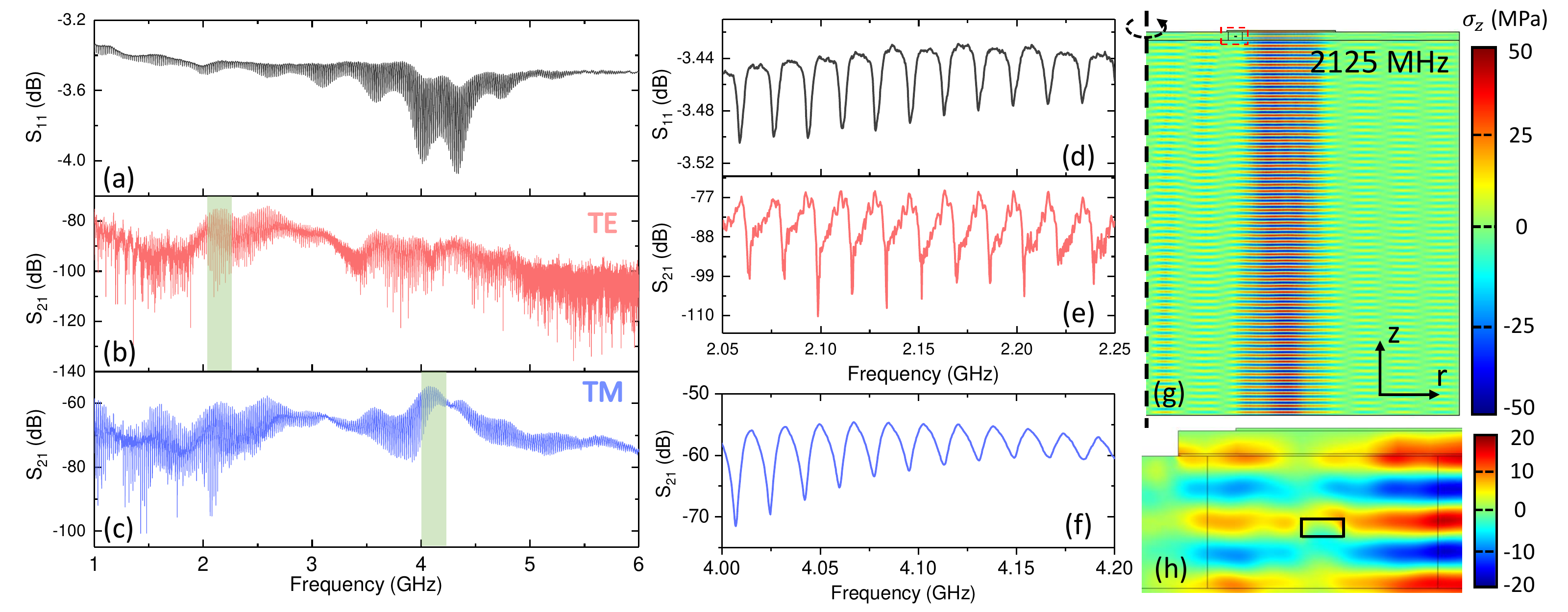}
\caption{The same S$_{11}$ and S$_{21}$ measurements on the Donut device in GHz range. (a) Electromechanical S$_{11}$ spectrum from 1 to 6 GHz. Optomechanical S$_{21}$ responses of (b) TE and (c) TM modes demonstrate effective stress-optical modulation spanning a broad range of microwave frequencies. The Donut device shows similar results as reported for the Disk device in the main text. (d) and (e) show the zoom-in of S$_{11}$ and S$_{21}$ responses of TE mode within the window (green shaded area) around 2 GHz in (b), while (f) shows the zoom-in of TM mode's S$_{21}$ response around 4 GHz in (c). (g) Numerical simulation of $\sigma _z$ distribution for one typical acoustic resonant mode at 2.125 GHz, with a zoom-in around the optical waveguide (red box) shown in (h). }
\label{Fig10}
\end{figure}

\subsection{RF frequency modulation of the Donut-shape device}
The same measurements described in the main text are also done for the Donut device, including S$_{11}$ and S$_{21}$ responses as shown in Fig. \ref{Fig10}. No big differences can be found between the Disk and Donut devices in terms of mode distribution, envelope of resonances, and signal to noise ratio. In the zoom-in around 2 GHz in Fig. \ref{Fig10}(e), there are multiple peaks inside each resonance due to existence of higher order acoustic modes. Numerical simulations of one of the fundamental modes at 2.125 GHz is shown in Fig. \ref{Fig10}(g), and the zoom-in around waveguide is in Fig. \ref{Fig10}(h). From the mode distribution, we can see the acoustic wave is effectively excited and confined under the actuator vertically with little diffraction angle. The good mode confinement guarantees low cross-talk between closely placed actuators.

\begin{figure}[htbp]
\centering\includegraphics[width=\textwidth]{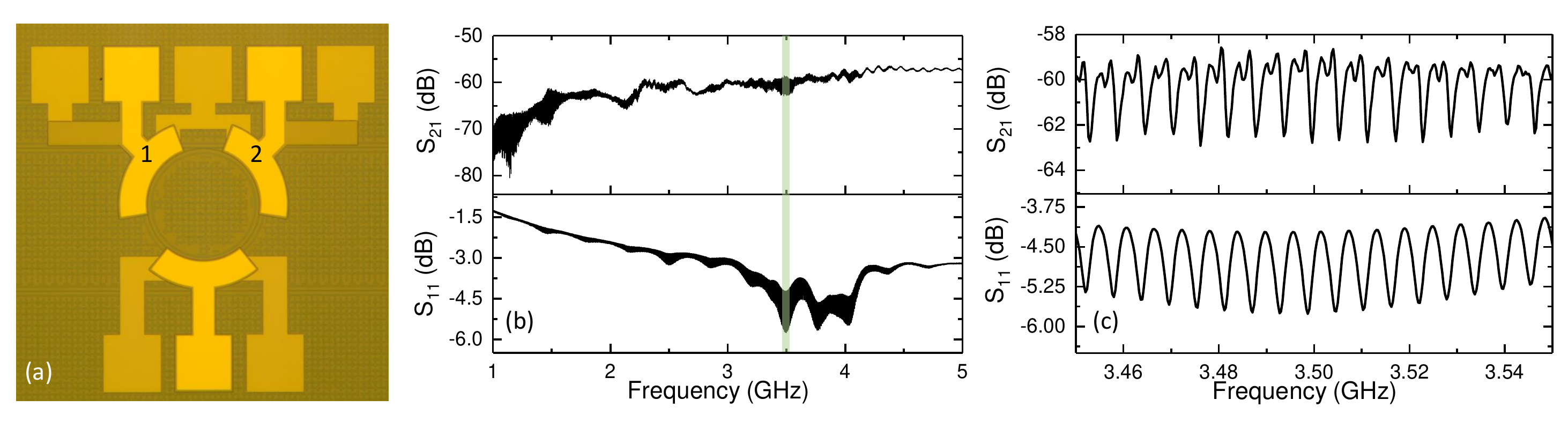}
\caption{Demonstration of low electromechanical cross-talk between adjacent actuators. (a) Optical microscope image of the device with three closely placed actuators. (b) (top) Two port electromechanical S$_{21}$ measurement by driving actuator 1 and sensing from actuator 2 as labeled in (a). (bottom) S$_{11}$ reflection for device 1. The cross-talk between the two devices is as low as -60 dB which guarantees compact integration. (c) Zoom-in of the measured S$_{21}$ and S$_{11}$ responses in the green shaded region in (b).}
\label{Fig11}
\end{figure}

\section{Electromechanical cross-talk between adjacent actuators}
As mentioned in the previous section, the high confinement of the acoustic mode beneath the actuator guarantees low cross-talk between adjacent actuators. To demonstrate this, three actuators are closely placed on the same optical ring resonator, which cover the whole ring in an interval of 120$^{\circ}$, as shown in Fig. \ref{Fig11}(a). The one port reflection parameter S$_{11}$ of actuator 1 is first measured as illustrated in Fig. \ref{Fig11}(b). The difference from the S$_{11}$ shown in the main text is caused by the thicker Si substrate (500 $\upmu$m) and thicker SiO$_2$ cladding (7 $\upmu$m). These lead to smaller FSR and period of the envelope. 

More importantly, the cross-talk is measured by performing a two-port electromechanical S$_{21}$ measurement, where we drive actuator 1 and sense the electrical signal out from the adjacent actuator 2. The cross-talk mainly comes from the leaking of acoustic waves from actuator 1 which can be sensed out by actuator 2 via the piezoelectric effect. The leakage of electric field will also be sensed by device 2 and cause cross-talk between electrical signals of device 1 and 2. As demonstrated in Fig. \ref{Fig11}(b), S$_{21}$ as low as -60 dB of cross-talk is achieved, which illustrates the electrical and mechanical isolation between the two adjacent devices. This low cross-talk enables us to fabricate several actuators on the same optical ring, which may realize optical isolation through spatial-temporal modulation \cite{shi2018, sohn2018time}, or dispersion engineering of Si$_3$N$_4$ microring resonator \cite{yao2018gate} by engineering stress distribution.

\clearpage
\spacing{1}

\section*{References}
\bigskip
\bibliographystyle{naturemag}
\bibliography{sample}
\end{document}